\documentclass[iicol, pdflatex, sn-basic]{sn-jnl}

\usepackage{lineno}
\usepackage{graphicx}
\usepackage{multirow}%
\usepackage{amsmath,amssymb,amsfonts}%
\usepackage{amsthm}%
\usepackage{mathrsfs}%
\usepackage[title]{appendix}%
\usepackage{xcolor}%
\usepackage{textcomp}%
\usepackage{manyfoot}%
\usepackage{booktabs}%
\usepackage{algorithm}%
\usepackage{algorithmicx}%
\usepackage{algpseudocode}%
\usepackage{listings}%

\begin{document}

\title{The ``Second Stellar Spectrum:''  Rotating hot massive star linear spectropolarimetry with the {\"O}hman Effect }

\author*[1]{J. Patrick Harrington}
\affil[1]{Department of Astronomy, University of Maryland, College Park, MD, USA}

\author*[2]{Richard Ignace}
\affil[2]{Department of Physics \& Astronomy,
East Tennessee State University,
Johnson City, TN 37614, USA}

\author*[3]{K.\ G.\ Gayley}
\affil[3]{Department of Physics \& Astronomy, University of Iowa, Iowa City, IA, USA}

\author*[4]{Jeremy J.\ Drake}
\affil[4]{Lockheed Martin Solar and Astrophysics Laboratory, 3251 Hanover St, Palo Alto, CA 94304, USA}

\abstract{
To understand better the polarized radiative transfer near the surface of rotating massive stars that remain nearly spherically symmetric, we use plane-parallel stellar atmosphere models to explore the unique opportunity presented by the \"{O}hman effect. This effect refers to the predicted variation in linear polarization across a rotationally broadened absorption line, due to the interaction of that line with the spatially varying continuum polarization across the face of a strongly scattering  photosphere, such as found in hot stars.  Even if the rotation is weak enough for the star to remain spherically symmetric, the \"{O}hman effect persists because differential absorption induced by the rotational Doppler shift of the line breaks the symmetry that would otherwise cancel the continuum polarization in the absence of that line.  Neglecting rotational distortion effects, the net polarization across the line vanishes, yet resolved line profiles display a telltale triple-peak polarization pattern, with one strong polarization peak at line center and two smaller ones in the line wings at a position angle that is rotated 90 degrees from the line center.  The far ultraviolet (FUV) is emphasized because both the polarization amplitude and the specific luminosity are greatest there for photospheres with effective temperatures between about 15,000 and 20,000~K.  Additionally, larger polarizations result for lower-gravity atmospheres.  There is a high density of spectral lines in the FUV, leading to a rich ``second stellar spectrum'' in linear polarization (analogous to the ``second solar spectrum'') that is made observable with stellar rotation.  Some hot stars exhibit extreme rotation, which suppresses the polarimetric amplitude for the forest of weaker FUV lines, but a few strong lines such as the Si{\sc iv} 140~nm doublet still give observable polarizations at high rotation speeds even before rotational distortion effects of the atmosphere are considered.  Thus polarizations at the level of 0.1\% to 1\% are achievable across individual lines for a wide variety of B-type stars.  
We highlight the prospects for accessing the unique information encoded in the \"{O}hman effect with future moderate-resolution spaceborne spectropolarimetric missions in the FUV.
}


\maketitle

\section{Introduction}  \label{intro}

Stars are generally not spherically symmetric, but the degree of symmetry breaking has a large dynamic range from the mild, such as starspots \citep[e.g.,][]{2002AN....323..213F, 2009A&ARv..17..251S, 2018MNRAS.473.5532R}, to the prominent, such as rotational distortion from rotating at near-critical rates \citep[e.g.,][]{2006Natur.440..896P, 2007Sci...317..342M, 2017NatAs...1..690C}.  Not only can the  star itself display aspherical features, so can its circumstellar environment, such as clumpy wind flows \citep{1991A&A...247..455H, 1994ApJ...421..310M, 2007A&A...476..335W} and co-rotating interaction regions  (CIRs) \citep{1984ApJ...283..303M, 1997A&A...327..699F, 2006JGRA..111.7S01T}.  Moreover, stars in close binaries can influence each other via tidal distortions \citep{1976ApJ...203..182W, 2014ARA&A..52..171O} and mass transfer \citep{2013ApJ...764..166D, 2017MNRAS.471.4256V}, or colliding winds \citep{1992ApJ...386..265S, 1997ApJ...475..786G}.  

For individual or close binary stars that are not spatially resolved, linear polarization measurements have a proven track record in stellar astrophysics as a powerful probe of geometry \citep{2010stpo.book.....C}.  Here we explore a diagnostic that has as yet to be observed called the \"{O}hman effect, which produces a polarimetric signature in the photospheric lines of rotating stars \citep{1946ApJ...104..460O}.  Although the principle applies widely to many types of stars, our focus will be on hot stars and their especially strongly scattering photospheres at far-ultraviolet (FUV) wavelengths, and the forest of FUV lines accessible to space-borne instruments \citep{2023MmSAI..94b.294N, 2023SPIE12676E..12B, 2024arXiv241001491M, 2024ApOpt..63.1710W, 2024BSRSL..93..156I, 2024sf2a.conf..457G, 2025arXiv250305556N, 2025arXiv250402659I}.

The continuum radiation emerging from a stellar atmosphere will in general be polarized by scattering, especially from free electrons. This
polarization varies with the angle between the emergent ray and the normal to the surface, and is greatest near the stellar limb. 
\cite{1946ApJ...103..351C, 1960ratr.book.....C} solved the radiative transfer equations including polarization for a plane-parallel atmosphere with opacity due entirely to scattering by a Rayleigh (dipole) phase function. This was expected to relate to hot stellar atmospheres where electron scattering can be the main opacity source. This solution is independent of wavelength and results in a maximum polarization at the limb of 11.7\%. 

Unfortunately, pure electron
scattering opacity is not realistic for hot stars. The appropriate equations
for a mixture of scattering and absorption were formulated by \cite{1950ApJ...112...22C}, who 
considered atmospheres with a constant ratio of scattering to absorption. He showed that the degree of polarization depended upon (i) the ratio of scattering to absorption, and (ii) the gradient of the source function with depth in the atmosphere. The gradient of the source function depends strongly on the wavelength.  In particular, for a given temperature gradient, the resulting gradient in the Planck function is much smaller at long wavelengths compared to wavelengths near the peak of the Planck function.  Consequently, the polarization from a hot star at visual wavelengths in the Rayleigh-Jeans tail of the emission will be far smaller
than Chandrasekhar's pure scattering result, but wavelengths in the ultraviolet will climb higher toward that level.

For the atmospheres of cool stars, \cite{1969ApL.....3..165H} suggested that there could be substantial polarization due to Rayleigh scattering by molecular
hydrogen.  At those lower temperatures, the gradient of the source function is steep at visual 
wavelengths, and his solutions showed that the polarization might actually 
exceed that of the Chandrasekhar pure scattering case.  But as with Chandrasekhar's original results, this early work was not based upon realistic model atmospheres.  

Modern atmosphere calculations include more complete physics over a full range of 
effective temperatures, gravities and chemical compositions.
However, this improved physics generally does not include the polarization, because the correct limb darkening and  unpolarized radiation intensity is satisfactorily achieved using simple isotropic scattering \citep{1970Ap&SS...8..227H}.  Thus it is expressly not designed to recover the correct polarization of a dipole phase function.

One reason the polarization from stellar atmospheres is often ignored is that even when the local limb polarization is large, the global symmetry causes the polarization to cancel when integrated across the unresolved face of a near-spherical star.  However, there are many situations in which spherical symmetry can be broken.  One example is close binaries, where there can be polarization from reflection that results in a net polarization from the anisotropic illumination of each star by the other \citep{2019NatAs...3..636B}.  For suitable orbital inclinations, eclipses can also lead to polarimetric variability \citep{1983ApJ...273L..85K}.  And for an individual star, extreme rotation can result in a net polarization through rotational distortion and gravity darkening effects \citep{1968ApJ...151.1051H}.  

This paper considers more common and more modest breaks from spherical symmetry, stemming solely from rotation that is not fast enough to distort the star, but does significantly broaden the photospheric absorption lines.  Such rotation produces a surface Doppler shift that allows each wavelength in the line to resonate with a different sector on the surface of the star, allowing the line to partially mask the continuum polarization in that sector, interdicting the global cancellation of the polarization at that wavelength.   This is the \"{O}hman effect, first described by \cite{1946ApJ...104..460O}.  

For a moderately rotating but otherwise spherically symmetric star, the isovelocity zones are vertical strips on the star that run parallel to the projected spin axis.  Considering a pure absorption line, at a given velocity shift in the line, $\Delta v$, flux will be depressed in the appropriately resonant spatial strip.  However, the other nonresonant strips contribute normally to continuum emission, and hence its polarization, allowing a net polarization to appear at that wavelength in the rotationally broadened line.  The line does not produce its own polarization, it merely blocks existing polarization, so in effect the line acts like a source of negative polarization, and that is what is observed.

The first-order result from the \"{O}hman effect is a symmetric line profile polarized either along or perpendicular to the rotational axis \citep{1946ApJ...104..460O}, which is termed the Stokes $Q$ flux and is conventionally positive for polarization along the axis and negative perpendicular to that.  This polarized profile occurs even for a spherical star, and in that symmetry, the total Stokes Q flux must integrate to zero across the line.  

A second but often weaker effect can appear in the Stokes $U$ flux profile, which refers to polarizations at 45 degrees to the $Q$ flux.  This signal indicates the presence of stellar distortion by near-critical rotation, otherwise the $U$ signal is null from geometric cancellation.  The $U$ polarization also vanishes at high inclination because of hemispheric cancellation, but at intermediate inclinations can produce a polarization   comparable to $Q$ \citep{2017NatAs...1..690C, 2022Ap&SS.367..124J, 2024ApJ...972..103B}.  High rotation rates can also lead to equatorial disks, especially in the early B stars considered here, which could complicate the interpretation of line diagnostics because the metal line polarization from disks has not been carefully examined to date.  Our primary focus is on low to modest rotation rates seen from high inclination, rather than very high rotation rates seen from moderate inclination, so the Stokes $U$ response of the \"{O}hman effect will be left to future more unified studies that include late B stars, and also consider the polarizing effects of disks.

The first deep analysis of the \"{O}hman effect in hot stars was by \cite{1991MNRAS.253..167C}, who presented an analytic treatment based on
 what they termed the ``Struve-Uns{\"o}ld model''.
 In that simplification, the rotating star is spherical and uniformly bright and its atmosphere purely scatters the upwelling radiative continuum, except for the presence of an absorption line treated in the limit of being intrinsically deep and narrow.  This resulted in an analytic 
expression for the line polarization peaking at line center and reaching an upper limit of $\sim 4\%$.  They then achieved a lower and more accurate estimate by replacing the uniformly bright star with a model that includes a parameterized limb darkening.
This analytic analysis demonstrated the basic workings of the {\"O}hman effect, and the qualitative shape of the polarization profile. 
They acknowledged that more quantitative results need to discard the pure
scattering model and consider the emergent intensity $I(\mu)$ and Stokes $Q(\mu)$
of a realistic model atmosphere, including how 
$I_\lambda(\mu)$ and $Q_\lambda(\mu)$ within the line differ from the simple continuum limb darkening. 

That paper continued with even more detailed corrections, applying the model atmospheres discussed in \cite{1991ApJS...77..541C},
which are based on the ATLAS6 model atmospheres. They considered a star of spectral type B1~V ($T_{\rm eff}\simeq 25,000K, \log g = 4.0$) and evaluated the effect in the FUV, using Si{\sc iii} 1113~\AA\ lines, and in the optical, using the He{\sc i} 4026~\AA\ line. For the Si{\sc iii} line, they found a maximum $Q/I$ polarization of 0.4\%, exhibiting the need for a complete treatment of the atmospheric gradients and absorption to avoid overestimating the polarization.
The reductions were much more severe in the optical domain, as the He{\sc i} line achieved only a miniscule peak polarization of $Q/I = 0.004$\%. 

These predicted reductions in the polarizations implied that the \"{O}hman effect could only be detected by combining high spectral resolution with excellent polarimetric sensitivity, and possibly only in the FUV.  
The prospects of spaceborne satellite missions have opened the possibility of a quantitative FUV \"{O}hman effect study in hot stars \citep[e.g.,][]{2023MmSAI..94b.294N, 2024sf2a.conf..457G,2024BSRSL..93..156I,  2024arXiv241001491M, 2025arXiv250305556N, 2025arXiv250402659I}.
The purpose of this paper is to further advance the quantitative study of this ``second stellar spectrum'', analogous to the ``second solar spectrum'' \citep{2002sss..book.....G}.  The latter describes the full Stokes IQUV spectrum of the Sun, which reveals a complex and rich set of diagnostics for scattering and absorption in the Sun's atmosphere along with influences from its magnetic fields.  In our case, assuming non-magnetic atmospheres, we anticipate a rich spectrum in linear polarization for spatially unresolved stars, in the form of characteristic 
\"{O}hman effect signatures from a forest of FUV lines.  Our goal is to capitalize on the wide range of rotation hot stars exhibit, to segment the line absorption across their polarized faces and extract the unique information that polarization encodes.

Section~\ref{sec2} describes numerical calculations of the \"{O}hman effect for hot stars.  It begins first by anticipating the polarization signature in a rotationally broadened line using a Milne-Eddington atmosphere.  Results for the ``second spectrum'' are presented for lines at FUV wavelengths with a focus on temperatures for B stars.  Then in Section \ref{sec3} we explore the fact that the high spectral density of line transitions in the FUV amounts to a statistical distribution of polarimetric fluctuations.  Further observational prospects are discussed in the context of a simulation based on the {\em Polstar} concept design for an FUV spectropolarimetry mission. Summary remarks are given in Section~\ref{sec4}.  Two appendices are provided.  The first in Section \ref{appA} gives technical details for evaluation the polarization from plane-paralle atmosphere models, and the second in Section \ref{appB} provides additional figures for a grid of polarized spectra in effective temperatures and surface gravities relevant for massive hot stars.

\section{Calculations for the \"{O}hman Effect} \label{sec2}

We draw on existing models for hot star atmospheres for our polarization studies.  This is possible from extracting both tabulated scattering and monochromatic continuum absorption as 
a function of depth\footnote{A grid of preliminary results for 52 hot stellar
models (15,000~K -- 50,000~K) and 438 cool star models (2500~K 
-- 6000~K) can be found at www.astro.umd.edu/$\sim$jph/Stellar\_Polarization.html.}.  In the case of hot stars, high temperatures ensure that the primary polarigenic opacity is free electrons. 
As a general rule, polarization is greater for stars of lower surface gravity due to their lower
density atmospheres \citep{2015A&A...575A..89K, 2016A&A...586A..87K}. 

A key point is that the amount of polarization depends on the anisotropy of the
radiation field near the surface (i.e., where $\tau \sim 1$), and this in turn depends on the  gradient of the source function. For hot stars at visible wavelengths, the radiation
field is not strongly peaked outward, and the polarization is found to be quite
small. In contrast, hot stars have strong polarization in the UV band because it is nearer the spectral peak.

\begin{figure}
\begin{center}
\includegraphics[width=\columnwidth]{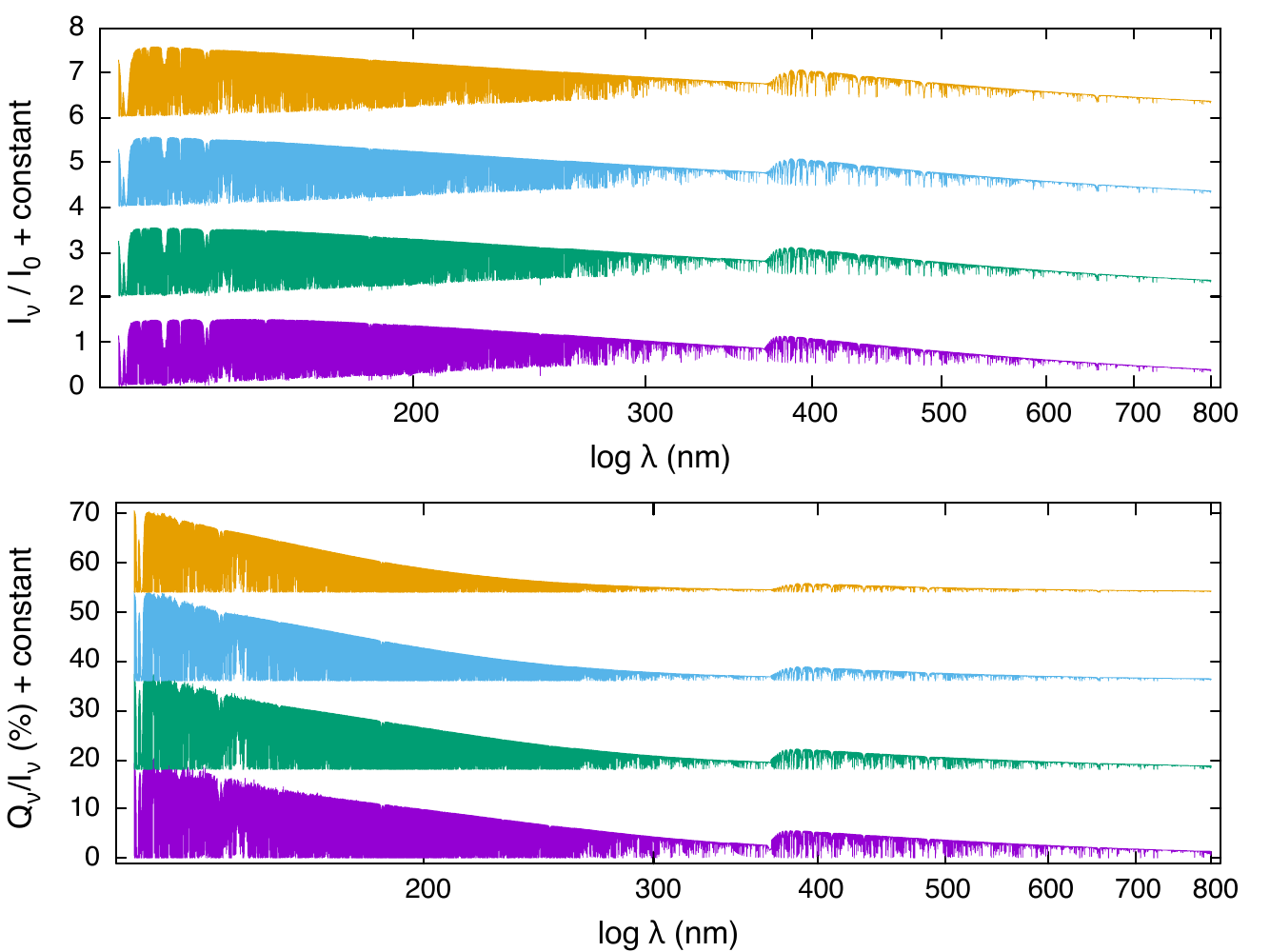}
\caption{The intensity distribution (top) and limb polarization (bottom) for stellar atmosphere models with $T_{\rm eff}=20,000$~K and $\log g = 3.0$ (purple), 3.5 (green), 4.0 (blue), and 4.5 (yellow).  These distributions are for very near the stellar limb at $\mu=0.005$.  Intensities are scaled and shifted for clarity of viewing.  The polarization is in percent and also shifted vertically for viewing.} 
\label{fig1}
\end{center}
\end{figure}

\begin{figure}
\begin{center}
\includegraphics[width=\columnwidth]{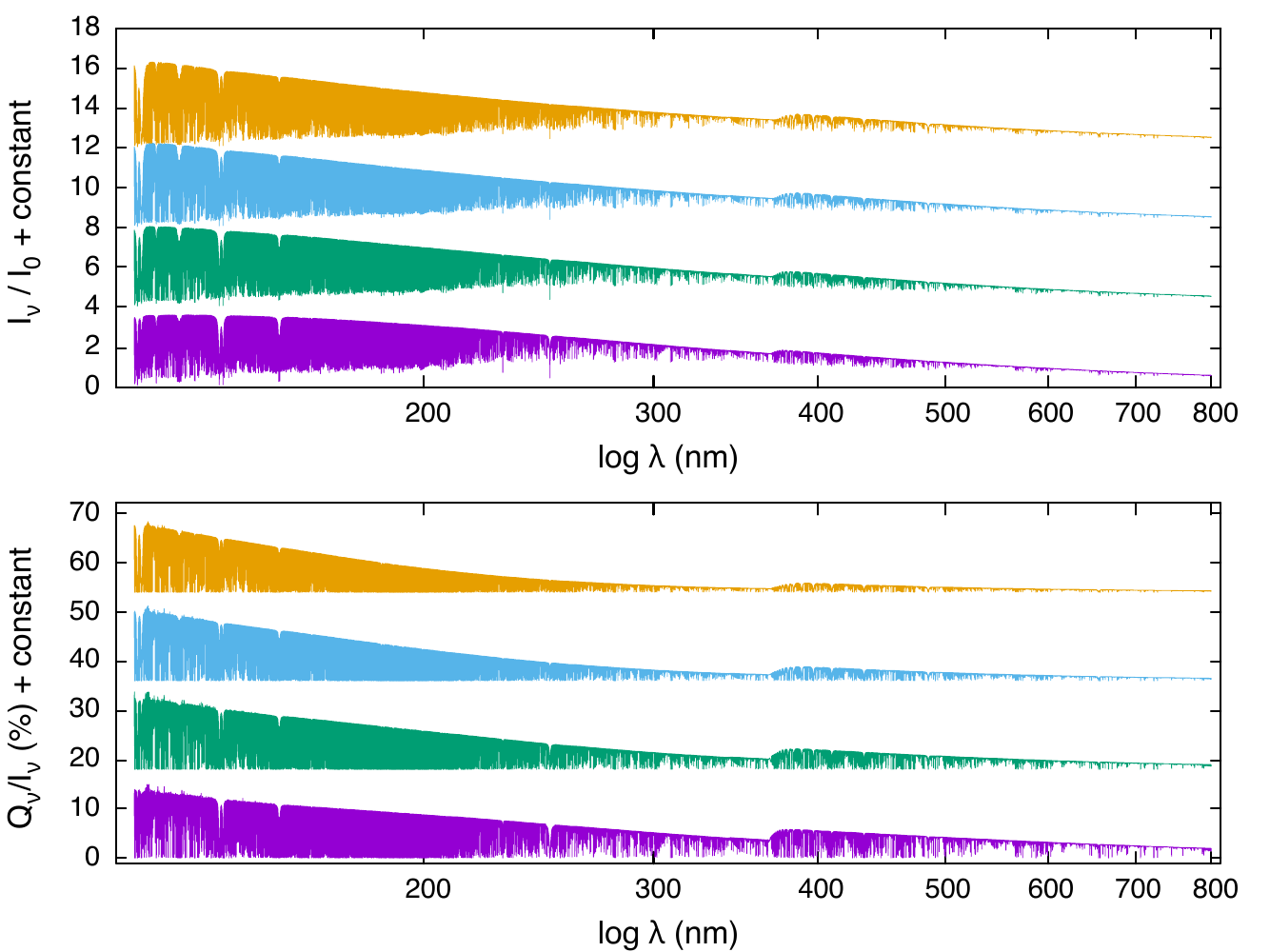}
\caption{As in Fig.~\ref{fig1} but now for $T_{\rm eff} = 25,000$~K.  The scale for polarization is the same, but different for intensity.} 
\label{fig2}
\end{center}
\end{figure}

\begin{figure}
\begin{center}
\includegraphics[width=\columnwidth]{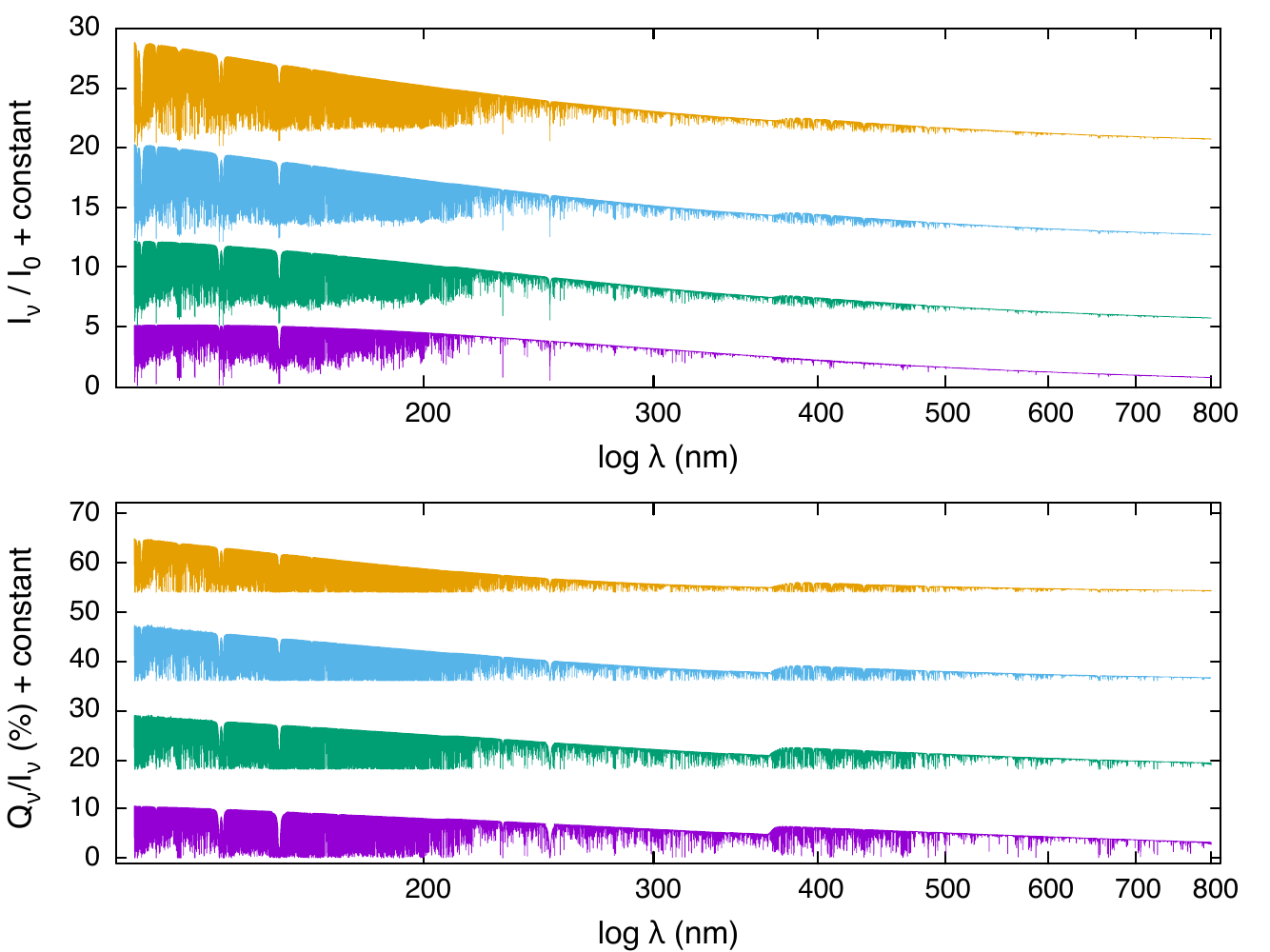}
\caption{As in Fig.~\ref{fig1} but now for $T_{\rm eff} = 30,000$~K.  The scale for polarization is the same, but different for intensity.} 
\label{fig3}
\end{center}
\end{figure}

Examples of the flux and polarization distributions for several hot stars are shown in Figures~\ref{fig1}--\ref{fig3}, for $T_{\rm eff}=20,000$, 25,000, and 30,000~K, respectively.  Purple, green, blue, and yellow are for $\log g = 3.0, 3.5, 4.0,$ and 4.5.  The fluxes in each top panel are conveniently scaled and shifted for ease of viewing.  The polarizations in each bottom panel are in percent; they are not scaled but are shifted for viewing.  The flux spectrum comes from integrating across the projected stellar annuli.
The polarizations are reported at $\mu = \cos \theta = 0.005$, near the limb.  Note that owing to spherical symmetry, the polarization would cancel identically with integration across the projected star.  Nevertheless, the plotted polarizations provide the back drop for calculation of net polarizations under conditions of symmetry-breaking.

These model spectra derive from TLUSTY atmosphere calculations \citep{2007ApJS..169...83L}.  The Stokes $I_\nu$ and $Q_\nu$ intensities as a function of $\mu$ were computed using the methods of \cite{2015IAUS..305..395H}, and detailed in Appendix~\ref{appA}.  It is these model atmospheres that will be employed for calculation of the \"{O}hman effect in rotationally broadened line profiles.  However, before presenting those results, the next section frames expectations using simpler considerations.

\subsection{``Milne-Eddington Lines'' in Model Atmospheres}


In the classical Milne-Eddington approximation \citep{1928MNRAS..89....3M, 1929MNRAS..89..620E}, the Planck function is taken as a linear function of optical depth and the ratio of line-to-continuum absorption is taken as constant with optical depth. These assumptions allow an analytic solution of the transfer equation. To achieve some additional quantitative accuracy, we will take the variation of the temperature and, most importantly, the variation of the ratio of scattering to continuous absorption, from the model atmospheres. 

To model the basic signature of the \"{O}hman effect, we extract
the run of temperature $T(\tau)$, continuum absorption coefficient $\kappa_{\rm c}(\tau)$ and continuum  
scattering coefficient $\sigma(\tau)$, as a function of monochromatic optical depth $\tau$, and use these to construct what we will call a ``Milne-Eddington'' line because we take the ratio of line to continuum absorption to be independent of depth.
Thus we replace the continuous absorption $\kappa_\nu^c(\tau)$ with

\begin{eqnarray}
\kappa_{\rm x}(\tau) & = & \kappa_\nu^c(\tau)\left[1~+~r_l~\phi(x)\right]~\mbox{and} \label{eq:kappa} \\
\phi(x) & = & \frac{1}{\sqrt{\pi}}~e^{-x^2}, \label{eq:linefunc}
\end{eqnarray}

\noindent where $r_l$ is the ratio of line-to-continuum absorption and $\tau$ is the optical depth in the continuum (absorbing plus scattering) at the line wavelength $\lambda_0$. Here we assume a Doppler line profile, where $x$ is the frequency in Doppler widths: $x=(\lambda -
\lambda_0)/\Delta \lambda_D$. The Doppler width, $\Delta \lambda_D$, can be written in terms of the thermal velocity $v_{\rm th}$ as $\Delta \lambda_D=(\lambda_0/c)~v_{\rm th}$, with $v_{\rm th}=12.85~[T/10^4 A]^{1/2}$ km/sec. Here, $A$ is the atomic weight of the atom in question, and for simplicity microturbulence is neglected. For stronger lines $\phi(x)$ is generalized as a Voigt profile. The parameter $r_l$ determines the line strength.

For each frequency $x$ across the line profile, the monochromatic
optical depth in the line, $\tau_{\rm x}$, is computed by integrating the opacity:

\begin{equation}
d\tau_{\rm x}(\tau)~=~\frac{\kappa_{\rm x}(\tau)~+~\sigma_\lambda(\tau)}{
         \kappa_\lambda^c(\tau)~+~\sigma_\lambda(\tau)}~d\tau 
\end{equation}

\noindent We thus obtain the Planck function $B_\lambda[T(\tau_{\rm x})]$ and the absorption to extinction ratio,

\begin{equation}
    \lambda_{\rm x} (\tau_{\rm x}) = \frac{\kappa_{\rm x}(\tau_{\rm {\rm x}})}{\kappa_{\rm x}(\tau_{\rm x})+
    \sigma_\lambda(\tau_{\rm x})}.
\end{equation} 

\noindent Interpolating these quantities for a pre-defined grid of 
optical depth and solving for $I_{\rm x}(\mu)$ and $Q_{\rm x}(\mu)$ and using finite integral transforms
as outlined in \cite{2015A&A...575A..89K} results in emergent Stokes intensities. Note that the limb darkening and Stokes $Q_{\rm x}$ will 
differ from the adjacent continuum behavior.  Usually, the added absorption
suppresses not only $Q_{\rm x}$ but the ratio $Q_{\rm x}/I_{\rm x}$ as well.  The signal is strongest when the line absorbers picked out by $x$ cover a strip of high positive or negative $Q_{\rm x}$, i.e., the center or edges of the stellar face, acting like a spatial filter that breaks the circular symmetry and leads to a net polarization at $x$.

Our treatment is limited to spherical stars.  The ways rapid rotation distorts the star's shape and redistributes its flux are not considered here.  These effects can be significant when centrifugal effects are strong enough to alter the effective gravity with colatitude $g = g(\theta)$, as in the von Zeipel approximation \citep{1924MNRAS..84..665V}, where the local effective surface temperature $T_{\rm eff}(\theta)
\propto g^{1/4}$.  This darkens the equator, while centrifugal effects also oblate its radius.  But these effects only become significant for quite rapid rotation, as even a star that is rotating as a solid body at half its equatorial break-up velocity will increase its 
equatorial radius, and decrease its equatorial temperature, by only around 4\%.  Such effects have been discussed by \cite{1991MNRAS.253..167C}.


A coordinate system is introduced with the rotation axis of the star as the $z$-axis of a spherical coordinate system.
Let $\theta$ be the angle measured from the $z$-axis (the co-latitude) and let $\phi$ be the azimuthal (longitudinal) angle measured about the spin axis. For solid body rotation, the azimuthal velocity of any point on the 
stellar surface is given by $v_\phi(\theta)~=~v_{\rm rot}~\sin\theta$, where $v_{\rm rot}$ is the rotational velocity of the stellar equator. The line-of-sight Doppler velocity shift seen by 
a distant observer is given by

\begin{equation}
v_{\rm obs}~=~v_{\rm rot}~\sin\theta~\sin\phi~\sin i,
\end{equation}

\noindent where $i$ is the angle between the rotation axis and the direction of the observer. This velocity produces a Doppler shift of $\Delta\lambda~=~\lambda~v_{\rm obs}/c~$. Thus the Doppler shift at a given point on the stellar surface can be written as

\begin{equation}
\Delta\lambda(\theta,\phi)~=~3.3\times 10^{-6}~\lambda~v_{\rm rot}~\sin 
    i~\sin\theta~\sin\phi~~,
    \label{eq:vobs}
\end{equation}

\noindent where $\lambda$ and $\Delta\lambda$ are in {\AA} units and $v_{\rm rot}$ is in km~s$^{-1}$. This is the shift that must be applied to the Stokes intensity parameters $I$ and 
$Q$ of the line profile at each point on the stellar surface. 

Equation~(\ref{eq:vobs}) indicates that the Doppler shift of every point is scaled by $\sin i$. A star with a rotational velocity of $v_{\rm rot}$ viewed at an inclination $i$ will look exactly like a star with rotational velocity $v_{\rm rot}'~=~v_{\rm rot}~\sin i$ seen at $i = 90^\circ$, as long as the star itself remains spherical.

Combining the polarizations from different patches on the stellar surface requires Mueller matrix rotations of each patch to a common reference axis \citep{2010stpo.book.....C}, which we take to be
the projection of the stellar rotation axis. Let the projection
of the local normal onto the plane of the sky make an angle $\xi$ with the
$z$-axis. The local emergent $Q(\mu)$ for that $\xi$ must be rotated through the angle $\xi$. In short, $Q(\mu)$ is defined locally along a projected radial from the center of the star and oriented at $\xi$.  The Mueller matrix rotation changes the local $Q$ polarization to take on new Stokes parameters $Q'$ and $U'$ for the reference axis.  The intensity $I$ is invariant upon rotation.

The equations for Stokes intensity parameters defined by the projected stellar spin axis are:

\begin{eqnarray}
I'(\mu) & = & I(\mu)~~, \label{eq:Irot} \\
Q'(\mu) & = & Q(\mu)~\cos(2\xi)~~, \label{eq:Qrot} \\
U'(\mu) & = & Q(\mu)~\sin(2\xi)~~. \label{eq:Urot}
\end{eqnarray}

\noindent Note that the total polarized intensity, $P'^2 = Q'^2 + U'^2 = Q^2 = P^2$, is preserved.
From considerations of spherical trigonometry, the angle $\xi~$ is given by 

\begin{equation}
    \tan\xi = \sin\phi~\tan\theta
\end{equation}

\noindent For calculation of the \"{O}hman effect with a spherical star, the net $U'$ will be zero.  This occurs because for each strip of constant $v_{\rm obs}$, a co-latitude with intensity $U'$ will have a corresponding value $-U'$ in the opposite hemisphere, and the two will cancel by symmetry.  However, this will not generally be true for rotationally distorted stars viewed at arbitrary inclinations, permitting diagnostics of rotational distortion and inclination not included here.

For a distant observer, the Stokes luminosities are given by:

\begin{equation}
L^I_{\rm x}~= 4\pi R^2\int_{-\pi/2}^{\pi/2} \int_0^\pi \mu I_{\rm x}(\theta,\mu)~
   \sin\theta~d\theta~d\phi~~
\end{equation}

\noindent and

\begin{equation}
L^Q_{\rm x}~= 4\pi R^2\int_{-\pi/2}^{\pi/2} \int_0^\pi \mu Q_{\rm x}(\theta,\mu)~\cos(2\xi)~
  \sin\theta~d\theta~d\phi~~.
\end{equation}

\noindent The value of $\mu(\theta,\phi)$ follows from 

\begin{equation}
    \mu=\sin\theta~\cos\phi.
\end{equation}
 
For a given combination of $T_{\rm eff}$ and $\log g$ for a model stellar atmosphere, values of $\kappa^c(\tau)$,
$\sigma(\tau)$ and $T(\tau)$ can be extracted.  Values are interpolated at the selected line wavelength 
$\lambda_0$.  Using equations (\ref{eq:kappa}) and 
(\ref{eq:linefunc}), values of $\tau(x_{\rm i})$ and $\lambda(x_{\rm i})$ are computed for  
a grid $x_{\rm i}$ across the (half-)line profile. 
The transfer equations for the emergent $I(x_{\rm i},\mu)$ and $Q(x_{\rm i},\mu)$ intensities are solved at 
these wavelengths for a grid of 18 values of $\mu_k$, adequate for interpolation in $\mu$.
We sample the visible surface of the star with a grid of 181 points in $\theta$ and 121 points in $\phi$. 
Interpolation is used to find the emergent $I(x_{\rm i},\theta,\phi)$ and 
$Q(x_{\rm i},\theta,\phi)$ profiles. For the selection of equatorial rotation speed $v_{\rm rot}$, Doppler shifts are applied and Stokes $I(x_{\rm i})$ and $Q(x_{\rm i})$ are properly weighted and summed across the stellar surface to obtain the Stokes luminosities.

\begin{figure}
\begin{center}
\includegraphics[width=\columnwidth]{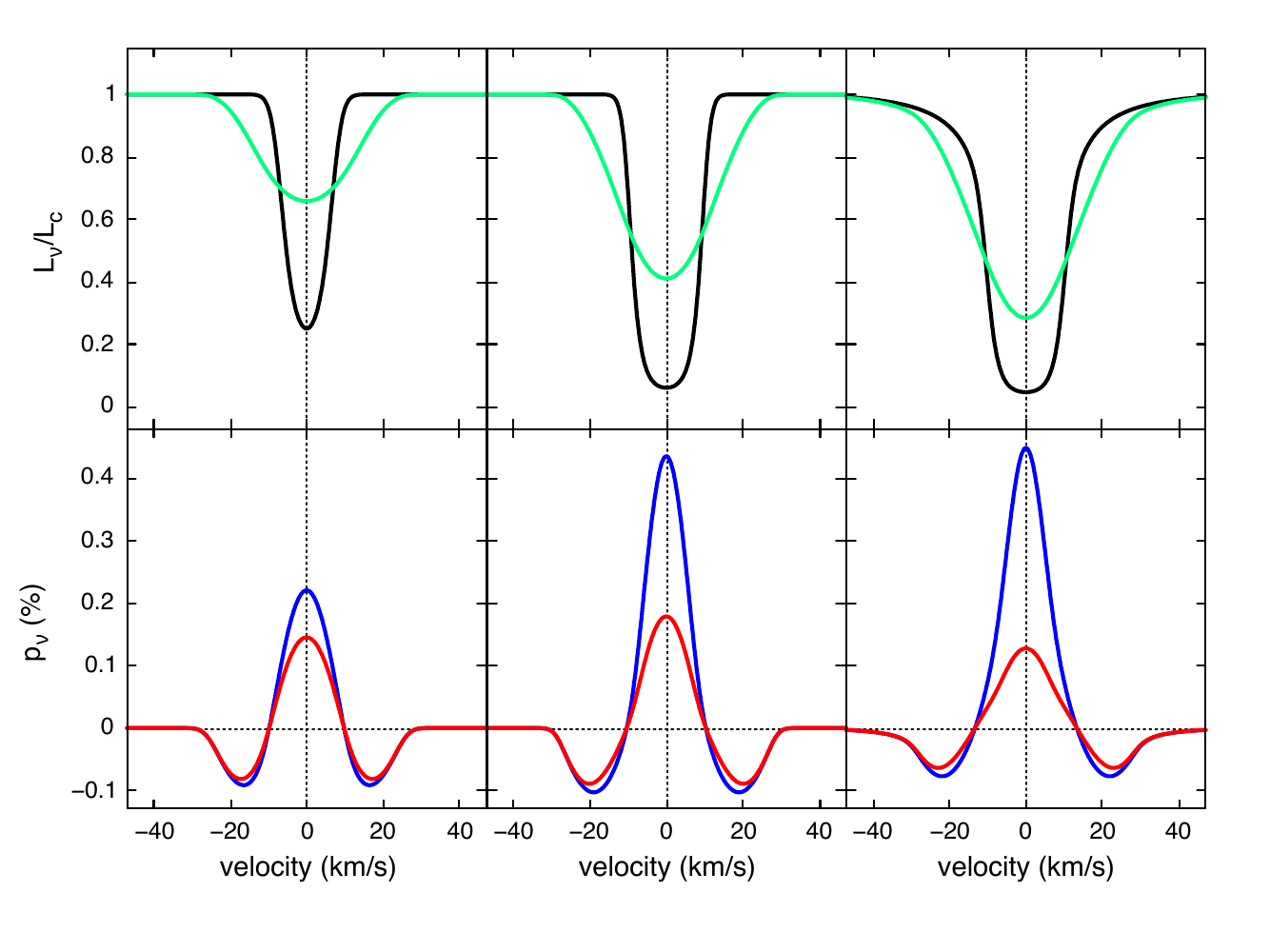}
\caption{The {\"O}hman effect using the ``Milne-Eddington'' approximation.  Upper panels show the continuum normalized lines in specific luminosity normalized.  The 3 lines are for $r_l = 10, 100$, and 200 (see text).  Lower panels show polarization $p_\nu$.  For these profiles the thermal speed is about 5~km/s assuming a hypothetical line for a carbon ion at 130~nm for a star of $T_{\rm eff}=17,000$~K.  The rotation is $4\times$ larger which maximizes the polarization amplitude.  At top black is the thermally broadened line, and green is for rotational broadening.  At bottom red is polarization for continuum normalization, whereas blue is normalized by the profile shape itself.  The former vanishes upon integration across the line whereas the latter does not.} 
\label{fig5}
\end{center}
\end{figure}

The resultant $Q_\nu (v_{\rm obs})$ curves involve not only  
the continuum limb darkening and the center-to-limb variation of the continuum polarization, but also how these interact with the shape of the line profile for a given rotational Doppler shift.  Three examples are displayed in Figure~\ref{fig5}.  The upper panels are continuum normalized absorption lines with $r_l=10$, 100, and 200.  The black curves are thermally broadened only.  The green curves are rotationally broadened at $v_{\rm rot}=4v_{\rm th}$ which maximizes the polarization signal.  Note that left and middle panels are for a Gaussian absorption profile, whereas the right panel is for a Voigt absorption profile.  The example calculation assumes a line from a carbon ion (i.e., $A=12$) at a fiducial wavelength of 130~nm for star of $T_{\rm eff} = 17,000$~K.  The thermal speed is about 5~km/s, so $v_{\rm rot} = 20$~km/s.  Note that the situation would be the same for any $v_{\rm rot} \sin i = 20$~km/s.

The lower panels are for the percent polarization across the lines.  Two curves are shown.  Red is $L^Q/L_{\rm c}$, which is just the Stokes $Q$ polarized luminosity profile normalized by the locally flat continuum.  This highlights the fact that integration across the line involving the two negative dips in the line wings (net polarization perpendicular to the spin axis) and the stronger central positive peak (net polarization parallel to the spin axis) will cancel to zero, as it must for spherical symmetry.  Observationally, it is standard to plot polarization as the ratio of polarized flux to the total flux, which will generally not integrate to zero over the line the way $Q$ must.  The total flux includes the line profile shape (green curves in the upper panels) which are the blue curves in the lower panels.  The wings are little affected by this normalization choice, while the central positive peak can be significantly enhanced, if the absorption line is deep.

From an observational perspective, the key conclusions are as follows.  The \"{O}hman effect leads to a signature triple-peak line profile in $Q$ linear polarization owing to the spatial filtering afforded by the Doppler broadening from rotation.  The polarized profile has a strong central peak with two weaker peaks of opposite sign in the line wings, where the switch in the sign corresponds to a polarization position angle rotation of $90^\circ$ from that of the central peak.  
Owing to these three extrema, the polarized profile concentrates more finely resolved structure into the same spectral width, compared to the singly dipped absorption line.

\begin{figure}
\begin{center}
\includegraphics[width=\columnwidth]{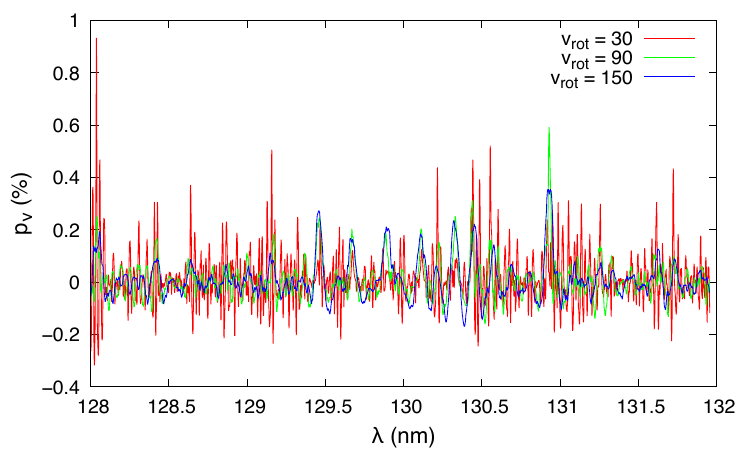}
\caption{A portion of the polarized spectrum for $T_{\rm eff}=15,000$~K with colors for the 3 rotation speeds as labeled.  The polarization amplitude generally decreases with faster rotation.} 
\label{fig6}
\end{center}
\end{figure}

\begin{figure}
\begin{center}
\includegraphics[width=\columnwidth]{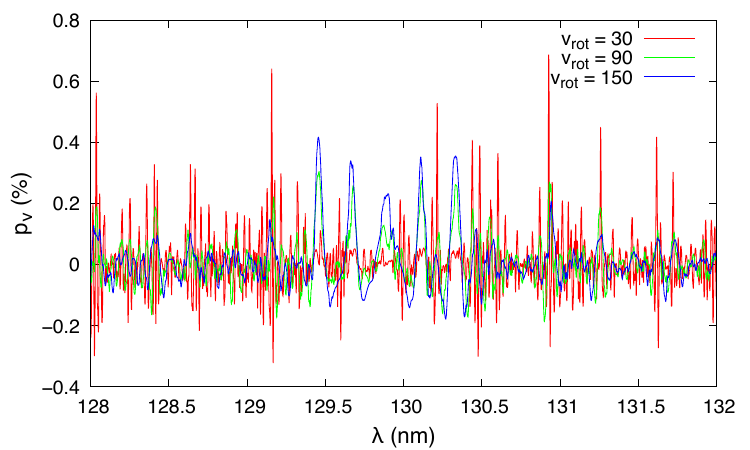}
\caption{As Fig.~\ref{fig6} but for $T_{\rm eff}=20,000$~K.} 
\label{fig7}
\end{center}
\end{figure}

\subsection{The ``Second Spectrum'' for Rotating Hot Stars}

Having established a general expectation from considerations of a single spectral
line, we now consider realistic spectra generated by the SYNSPEC (v. 49) program, which
makes use of stellar atmosphere models generated by the TLUSTY code
\citep[e.g.,][]{hubeny2011}.
While SYNSPEC generates spectra with thousands of lines, the output is without
polarization. It was therefore necessary to modify the SYNSPEC source code to
to write out, for each wavelength point, the line and continuum opacity, as well
as the scattering, for all tabulated depths in the atmosphere. This data enables
us to compute (as outlined in Appendix A) the Stokes parameters $I(\mu)$  and
$Q(\mu)$ of the emergent radiation. This then allows us to evaluate for each point
on the stellar surface, the polarized radiation for the relevant  $\mu$ and $\xi$.
We then apply the Doppler shift appropriate to each surface point and interpolate
back to the common wavelength grid. The result is $L^I_\nu$ and $L^Q_\nu$ with
$L^U_\nu=0$.

Appendix~\ref{appB} shows the results for a grid of hot star atmospheres, for combinations of $T_{\rm eff} = 15,000$, 17,000, 20,000, 25,000, and 30,000~K along with $\log g = 3.0$, 3.5, 4.0, and 4.5 using two rotation speeds of $v_{\rm rot} = 60$ and 120 km/s.  For reference, we use Figure~1 of \cite{2007ApJS..169...83L} to relate these combinations to estimates for fundamental stellar properties given in Table~\ref{tab1}.  The figure displays model isochrones for masses $M_\ast$ that span the grid of temperatures and gravities.  With those masses, we obtain radii from $\log g$ values, the critical speeds of rotational break-up $v_{\rm c}$, and finally bolobmetric luminosities $L_\ast$.  The properties are for solar abundances and tied to the isochrones adopted by in \cite{2007ApJS..169...83L}.

The overall trend of figures in Appendix~\ref{appB} is for low polarization at visible wavelengths so that only the FUV band is plotted.  Even there, the polarization amplitudes rarely rise above 1\%.  There is an evident trend of higher overall polarizations for lower gravities, and lower values for the hottest stars.  In fact for $T_{\rm eff}$ of 20,000~K and higher, spectra for $\log g = 4.5$ both because the polarization is small and because the gravities correspond to stars below the main sequence.  

\begin{table}
\caption{Stellar Properties for Select Model Atmospheres}   \label{tab1}
\begin{tabular}{cccccc}
\hline\hline $T$ & $\log g$ & $M_\ast$ & $R_\ast$ & $v_{\rm c}$ & $L_\ast$ \\
 (K) & (cm/s$^2$) & $(M_\odot)$ & $(R_\odot)$ & (km/s) & $(L_\odot)$ \\ \hline
 & & & & & \\
15,000 &	4.0 & 	4.5 &	3.5 &	490 &	560 \\
17,000 &	4.0 & 	5.5 &	3.9 &	520 &	1,140 \\
20,000 &	4.0 & 	7.0 &	4.4 &	550 &	2,780 \\
25,000 &	4.0 & 	11 &	5.5 &	620 &	10,600 \\
30,000 &	4.0 &	15 &	6.4 &	670 &	29,700 \\
 & & & & & \\
15,000 &	3.5 & 	5.5 &	6.9 &	390 &	2,160 \\
17,000 &	3.5 & 	7.0 &	7.8 &	410 &	4,540 \\
20,000 &	3.5 &	9.0 &	8.9 &	440 &	11,000 \\
25,000 &	3.5 &	15 &	11 &	510 &	42,000 \\
30,000 &	3.5 &	25 &	15 &	570 &	159,000 \\ 
& & & & & \\ \hline
\end{tabular}
\end{table}

The lower rotation speed produces generally higher polarization amplitudes.  This is an important point for practical issues of detectability.  Line broadening at the level of 60~km/s implies FWHM values of around 100~km/s.  An isolated absorption line shows a single absorption trough, but the \"{O}hman effect in polarization shows 3 peaks with 2 reversals:  a negative peak at negative velocity shifts from line center, a positive peak at line center, an another negative peak now at positive velocity shifts.  A sufficient resolving power $R$ is necessary to detect this profile morphologies.

\subsection{Comparison of Second Spectra between Early and Late B Stars}

For spherical hot stars, limb polarization is suppressed at visible and longer wavelengths because in the Rayleigh-Jeans limits, the source function gradient is reduced.  Limb polarization is higher around the region of the Wien peak, which for hot stars moves into the FUV.  As a practical matter, spectropolarimetric mission concepts for space-borne facilities are likely to focus on wavelengths longward of the Lyman limit.  As demonstrated with the {\em EUVE} mission, only two massive stars were observable because they were both relatively close and in the direction of low interstellar hydrogen column density \citep{1995ApJ...438..932C, 1996ApJ...460..949C}.  Additionally, stars hotter than 30,000~K move their Wien peaks below about 1000~\AA, whereas current mission concepts are focused longward of about 1200~\AA.  As a result, quite hot stars are essentially Rayleigh-Jeans by this wavelength, with lower continuum polarization levels as indicated in Figure~\ref{fig3}.

This section focuses on exploring detailed line effects for stars of 15,000~K and 20,000~K.  Figure~\ref{fig12} shows the stellar flux is starting to drop by 1200~\AA\ at around 15,000~K, and Figure~\ref{fig2} indicates that limb polarization levels are dropping by 25,000~K.  So stars between 15,000 and 20,000~K are likely ideal for exploring the Second Stellar Spectra in relation to plans for future missions.  Such temperatures roughly span the B~star spectral types for main sequence, subgiant, and giant luminosity classes.  Since bright giants and supergiants are generally quite slow rotators, and relatively rare, we consider $\log g = 3.5$ as representative.

In general, the {\"O}hman effect is most pronounced when the equatorial rotational velocity $v_{\rm rot}\sim 3v_{\rm th}$ for Doppler profiles and $\sim 15v_{\rm th}$
for Voigt profiles. The rotation speeds of interest for \"{O}hman range

\begin{equation}
v_{\rm rot}~\simeq~(40-200)~\left[\frac{T}{10^4 A}\right]^{1/2}~(\sin i)^{-1}~~~~\mbox{km/sec}~.
\end{equation}

\noindent For 15,000 and 20,000~K, and for edge-on, this corresponds to lower rotations of about 60 and 80 km/s for Doppler profiles, but 300 and 400 km/s for Voigt profiles.  

\begin{figure}
\begin{center}
\includegraphics[width=\columnwidth]{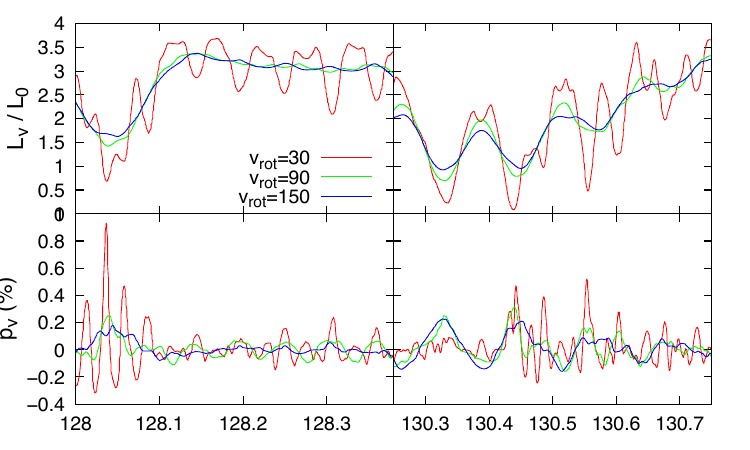}
\caption{For models with $T_{\rm eff}=15,000$~K, now zooming into a small interval of wavelengths for the shapes of individual lines can be seen in detail.  Top panels are relative fluxes; bottom are percent polarizations. Right and left are different wavelengths that highlight strong polarization values. The 3 colors are for the rotation velocities as labeled.}
\label{fig8}
\end{center}
\end{figure}

\begin{figure}
\begin{center}
\includegraphics[width=\columnwidth]{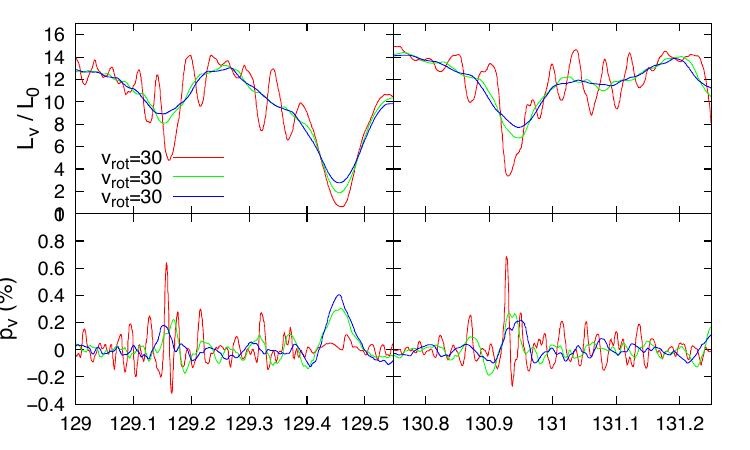}
\caption{As Fig.~\ref{fig8} but for $T_{\rm eff}=20,000$~K.} 
\label{fig9}
\end{center}
\end{figure}

Figures~\ref{fig6} and \ref{fig7} show the polarization spectra for 5 rotation speeds at 30, 90, and 150 km/s, as indicated, for wavelengths of 1280-1320~\AA.  The first is for 15,000~K and the second for 20,000~K.  Signals are largest at the slowest speed (purple), and amplitudes generally drop with more rotational broadening. However, a few lines are strongest at the highest speed.

Figure~\ref{fig8} for 15,000~K and Figure~\ref{fig9} for 20,000~K are selected portions of the FUV at 5~\AA\ widths to further zoom into the shapes of the polarized profiles (bottom panels) and also showing the Stokes luminosities as well (top panels).  The rotation speed values and associated colors are the same as in Figures~\ref{fig6} and \ref{fig7}.  The Stokes-I spectra are scaled and relative for ease of viewing; however, the polarization spectra all have the same vertical scale for Figures~\ref{fig6}-\ref{fig9}.

\begin{figure}
\begin{center}
\includegraphics[width=\columnwidth]{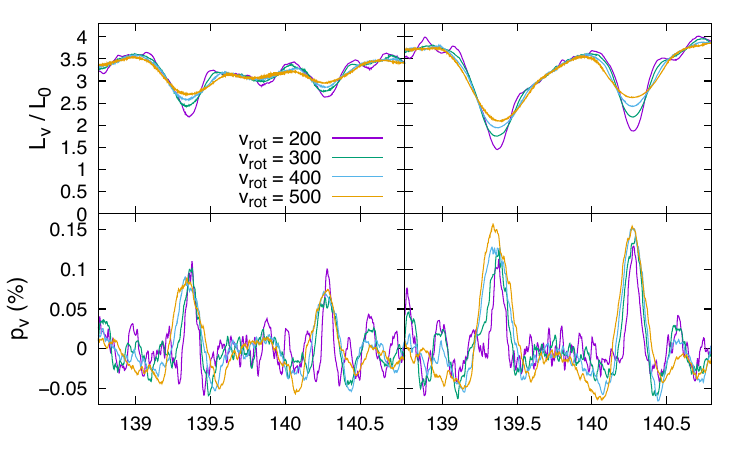}
\caption{Relative luminosity and polarization panels like the previous figure, now emphasizing the region around the Si{\sc iv} doublet.  The left panels are for $T_{\rm eff}=15,000$~K and the right panels for $T_{\rm eff}=20,000$~K.  The rotation speeds being explored are much higher.  No rotational distortion or gravity darkening effects are included; the star remains spherical.  The example highlights that strong lines like the Si{\sc iv} doublet will tend to retain a stable polarimetric signal even at high rotation speeds that tend to wash out weaker lines.} 
\label{fig10}
\end{center}
\end{figure}

\begin{figure}[h]
\begin{center}
\includegraphics[width=\columnwidth]{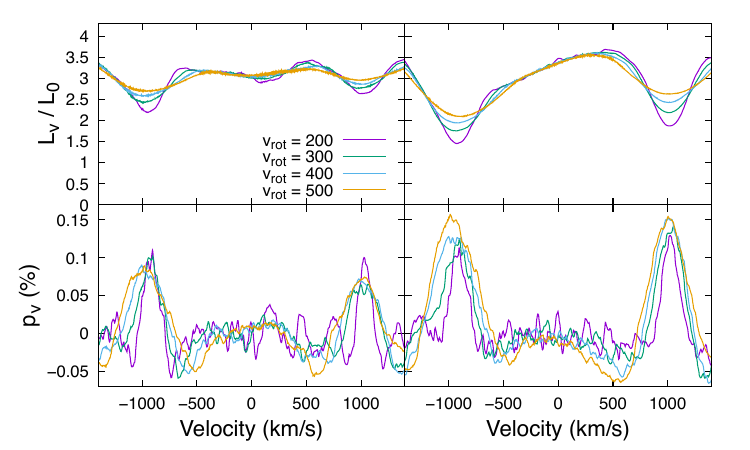}
\caption{This is Fig.~\ref{fig10} replotted in velocity shift units using 1398~\AA\ as the zero point.  Colors as in Fig.~\ref{fig10}.} 
\label{fig11}
\end{center}
\end{figure}

Figures~\ref{fig10} and \ref{fig11} focus on the Si{\sc iv} doublet which is described with a Voigt profile.  Figure~\ref{fig10} is plotted in wavelength, and Figure~\ref{fig11} is in velocity units, with zero set at 1398~\AA\ between the doublet components.  For each figure the two left panels are for 15,000~K, and the two right panels are for 20,000~K.  Instead of low rotation speeds, quite high rotations speeds of 200--500 km/s are used.

There are two key features.  First, by virtue of being at somewhat longer wavelength at 140~nm, the overall scale of polarization is lower at around 0.1\% as compared to lines around 130~nm.  Second, high rotation speeds strongly suppress the \"{O}hman effect in the numerous weak lines;  however, the strong doublet components of Si{\sc iv} remain relatively prominent.  Indeed as the speed increases, the weaker lines are strongly blended while the overall signature triple-peak of the \"{O}hman effect becomes more clear in the doublet component.  Interestingly, the polarization contrast from the two negative troughs to the central positive peak actually increases slightly.

\section{Observational Considerations for the Second Stellar Spectrum}   \label{sec3}

Here we explore two types of observations.  The first involves treating the spectrometric variations statistically in terms of polarimetric fluctuations owing to the high density of lines in the FUV.  In the second case, a simulation is made in the case of the {\em Polstar} mission concept for a future FUV spectropolarimetric spaceborne telescope.

\subsection{Polarimetric Fluctuations}

Figures~\ref{fig6}--\ref{fig9} of earlier sections provide narrower views for select intervals of wavelengths to highlight the \"{O}hman effect across strong and weak lines, effects for blends, and the effect of changing the stellar rotation speed.  However, what is remarkable about the broad wavelength view in the FUV is the overall statistical appearance of a choppy and even spikey Second Spectrum for polarization.  The polarimetric fluctuations depend on temperature for a fixed waveband, on the rotational broadening, and on stellar gravity. 

\begin{figure*}
\begin{center}
\includegraphics[width=0.85\textwidth]{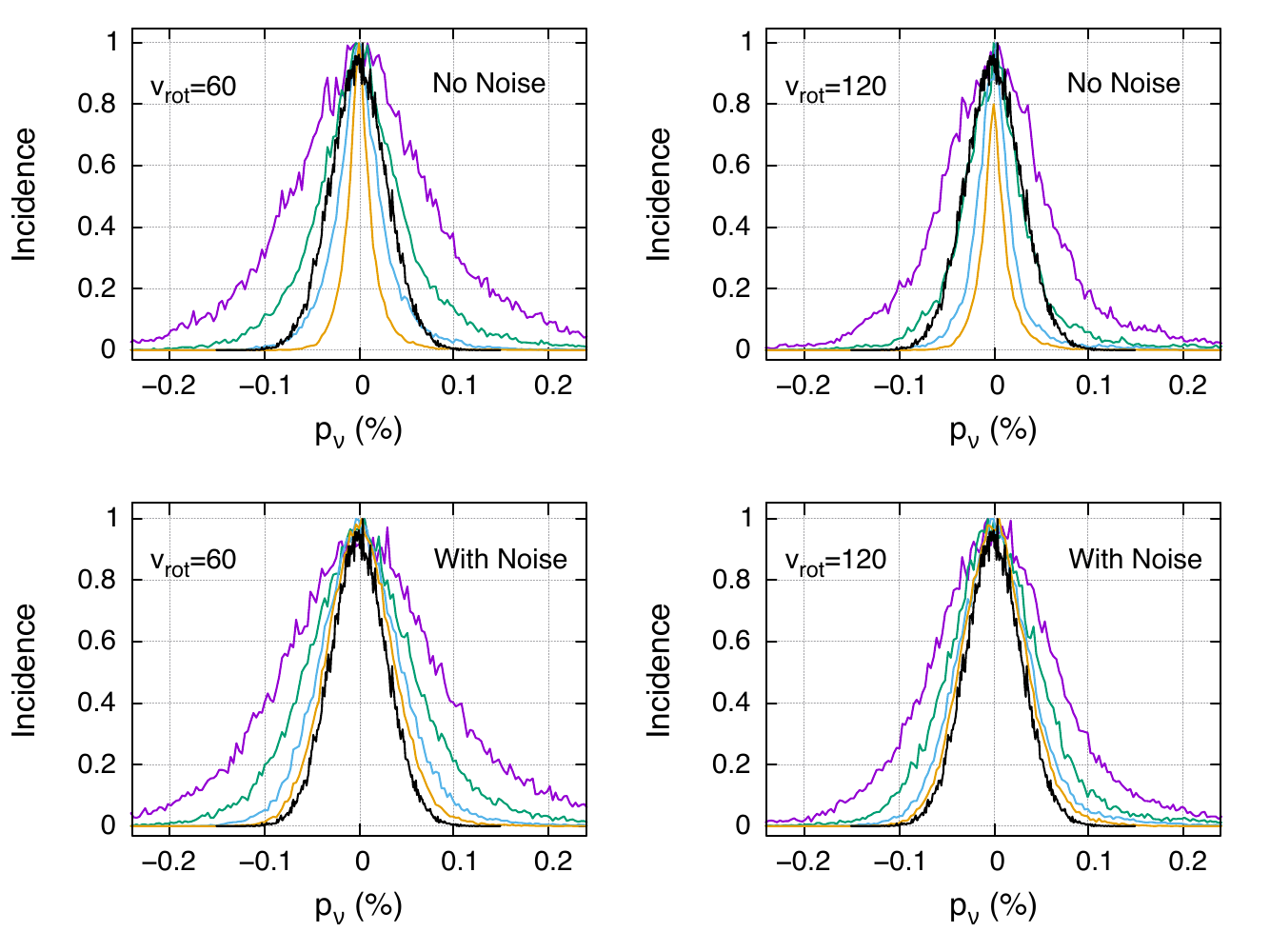}
\caption{A comparison of the polarimetric fluctuations across the wavelengths shown in Fig.~\ref{fig13}.  Colors correspond to those of that Figure for the respective values of $\log g$.  Left and right are $v_{\rm rot}=60$ and 120 km/s.  Top is the atmosphere model whereas bottom has noise added to simulate measurement uncertainties.  The distributions are normalized to a peak of unity.  The widths are wider for lower gravities.  At bottom, with Gaussian distributed noise at the level of 0.03\%.  The distribution of noise is indicated by the black curve in each panel.  The higher gravity atmospheres are dominated by the noise level, as demonstrated by (a) broadened distributions and (b) that the distributions with $\log g$ of 4.0 and 4.5 are nearly the same as the noise distribution.  The lower gravities of 3.0 and 3.5 are less affected for $v_{\rm rot}=60$ km/s but somewhat more affected for the higher rotation of 120 km/s where the distributions without noise added are intrinsically more narrow to begin with.} 
\label{fig17}
\end{center}
\end{figure*}

To characterize the statistical fluctuations from a broad interval of wavelengths, Figure~\ref{fig17} displays distributions for the polarization from about 120 to 165 nm.  For illustration purposes we have chosen the models for $T_{\rm eff} = 17,000$~K; other temperature would be similar in form but have a different spans in polarization values.  The upper panels are for the atmosphere models; lower panels have noise added to replicate the observational situation with finite SNR.  Left panels are for $v_{\rm rot}=60$ km/s and right for $v_{\rm rot} = 120$ km/s.  The colors are for the $\log g$ values appearing in Fig.~\ref{fig13}.

All the distributions are normalized to a peak of unity.  Clearly, lower gravity atmospheres have wider distributions.  Also, the distributions are asymmetric about zero.  Not shown is an extended wing for higher positive values of polarization as compared to the wing for negative values.  Recall that the signature \"{O}hman effect for an isolated line is two negative peaks and a stronger central positive one, hence the asymmetry and disparity in the wings of the distributions.

To simulate the impacts of limited SNR, the lower panels show distributions when Gaussian distributed noise with $\sigma_P=0.03\%$ for polarization precision is added.  The results are as expected.  For distributions with small physically intrinsic widths like $\log g$ of 4.0 adn 4.5, noise at this level for these models dominates, but the wings remain intact.  For lower gravities, 3.0 and 3.5 in the long, the intrinsic fluctuations dominate the noise.  

\subsection{Simulation of the \"{O}hman Effect for {\em Polstar}}

A variety of spectropolarimetric missions and/or instrument designs are in advanced planning stages.  Several of these were noted in Section~\ref{intro}.  One of these is the {\em Polstar} concept being developed for consideration as a NASA SMEX\footnote{See https://explorers.gsfc.nasa.gov for a description of the Explorers program.} mission.  {\em Polstar} is described in several articles noted in the Introduction and elsewhere in this special issue of the journal.  In summary, {\em Polstar} is planned as a 40~cm aperture telescope providing for measurement of all four Stokes parameters, IQUV.  Spectropolarimetry would be obtained at $R=20,000$ over the waveband of 121--285~nm.

\begin{figure}
\begin{center}
\includegraphics[width=\columnwidth]{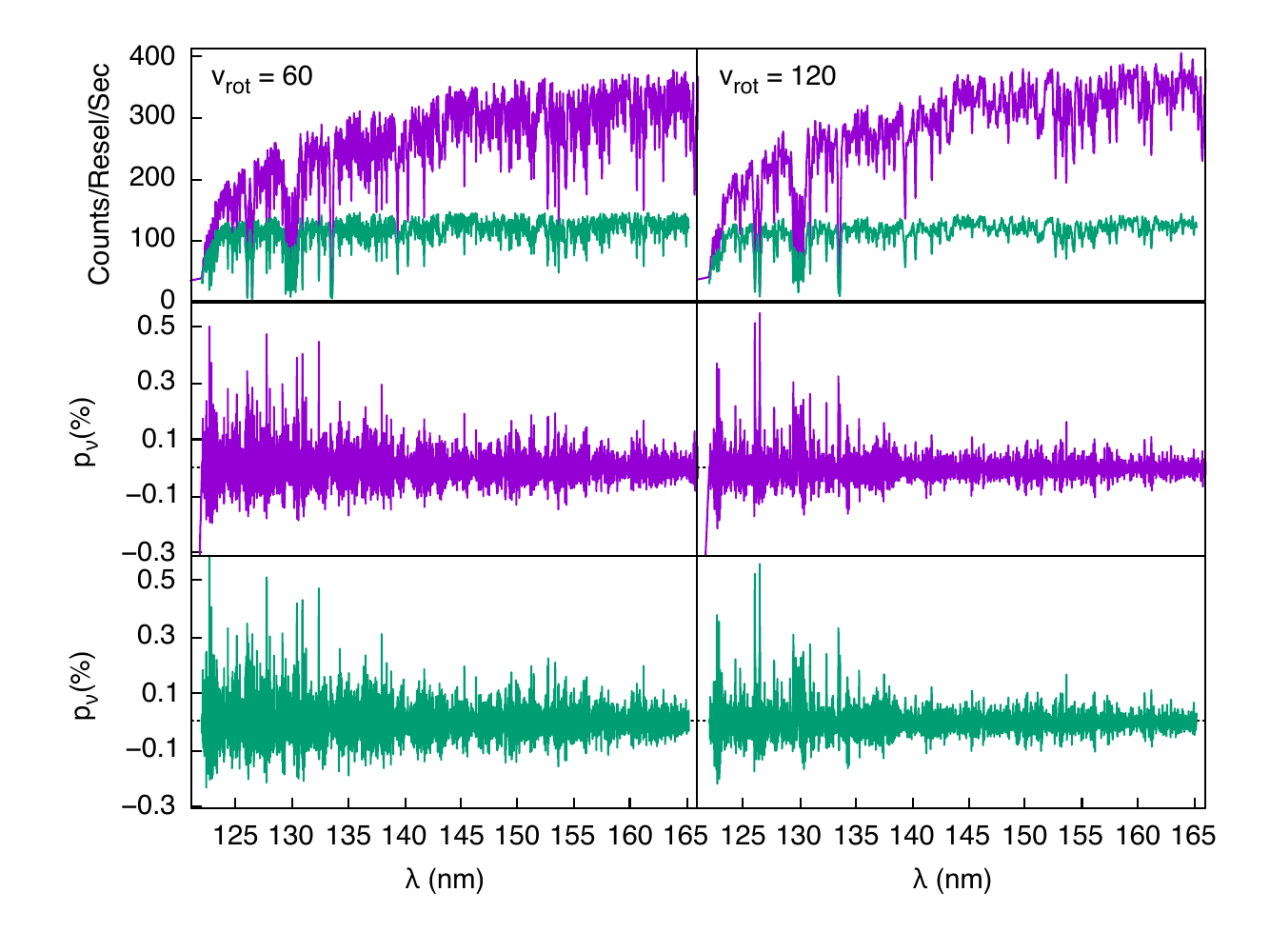}
\caption{Comparison of input spectra with a simulation based on the instrument design for {\em Polstar} (see text).  Left and right are for two different equatorial rotation speeds as labeled.  Purple is for the spectral simulation, whereas green is the input spectra.  Top is the simulation in counts/resel/sec, where ``resel'' refers toa  spectral binwidth at $R=20,000$.  The green spectrum is scaled for ease of viewing in comparision.  Middle panels are the input polarization spectra; bottom is the result of the simulation.} 
\label{fig18}
\end{center}
\end{figure}

Predicted spectra are displayed in Figure~\ref{fig18} using a software model of the throughput of {\em Polstar} that captures the collecting area and efficiencies of the various instrument subassemblies.  This includes the telescope collecting area and support structure obscuration, the reflectivities of MgF$_2$-coated primary and secondary mirrors, the entrance aperture fractional throughput, the transmission of the polarimeter modulator and Wollaston prism analyzer, the collimator reflectivity, the efficiencies of echelle and cross-dispersing gratings, and the quantum efficiency of the CMOS detector. The modulator comprises 4 plates of MgF$_2$ crystal, each approximately 0.3~mm thick, that are optically contacted such that there are no gaps between them. The Wollaston prism comprises two optically-contacted MgF$_2$ crystals with a total thickness of approximately 5~mm. The spectrum simulator includes instrumental broadening, although for the rotation velocities considered here instrumental broadening is small. 

For purposes of illustration, an input spectral model with a temperature of 17000~K and a surface gravity of $\log g=3.5$ were used. The model was first normalized to a flux density of $10^{-9}$~erg~cm$^{-2}$~s$^{-1}$~\AA$^-1$ at 150~nm, which is a value representing a reasonably bright source for the provisional {\em Polstar} target list. No interstellar extinction is included since the spectrum is scaled to a fiducial flux.  We did compute a model with some reddening which does alter the spectral slope across the waveband, but reddening is not included for this example.

Figure~\ref{fig18} shows 6 panels.  On the left are models with a stellar rotation of $v_{\rm rot}=60$ km/s; the right is for $v_{\rm rot}=120$ km/s.  Top panels show a comparison of the input spectrum in green and scaled for display purposes with the simulation for {\em Polstar} in purple.  The simulated spectrum is displayed as counts/resel/sec.  Here ``resel'' refers to a spectral binwidth corresponding to $R=20,000$.  Given that expected polarizations are generally small, of order 0.1--3\%, targets for {\em Polstar} will generally be bright and observed in the limit of shot noise.  Consequently the unit of display enables easy estimates of on-target exposure to achieve a requisite signal-to-noise ratio (SNR).

The middle and bottom panels shows polarization in percent.  Again, purple is for the simulation and green is for the input model.  Note that the input spectrum in green has a resolution of order $10^{5.2}$, several times larger than for {\em Polstar}.  However, the two polarization spectra look very similar because the rotational broadening is already reasonably resolved.  The main differences arise from line blends.  Because the polarization is signed (positive and negative), the lower resolving power of the simulation relative to the input sometimes leads to slight cancellations, but only for slower rotations.

\begin{figure}
\begin{center}
\includegraphics[width=\columnwidth]{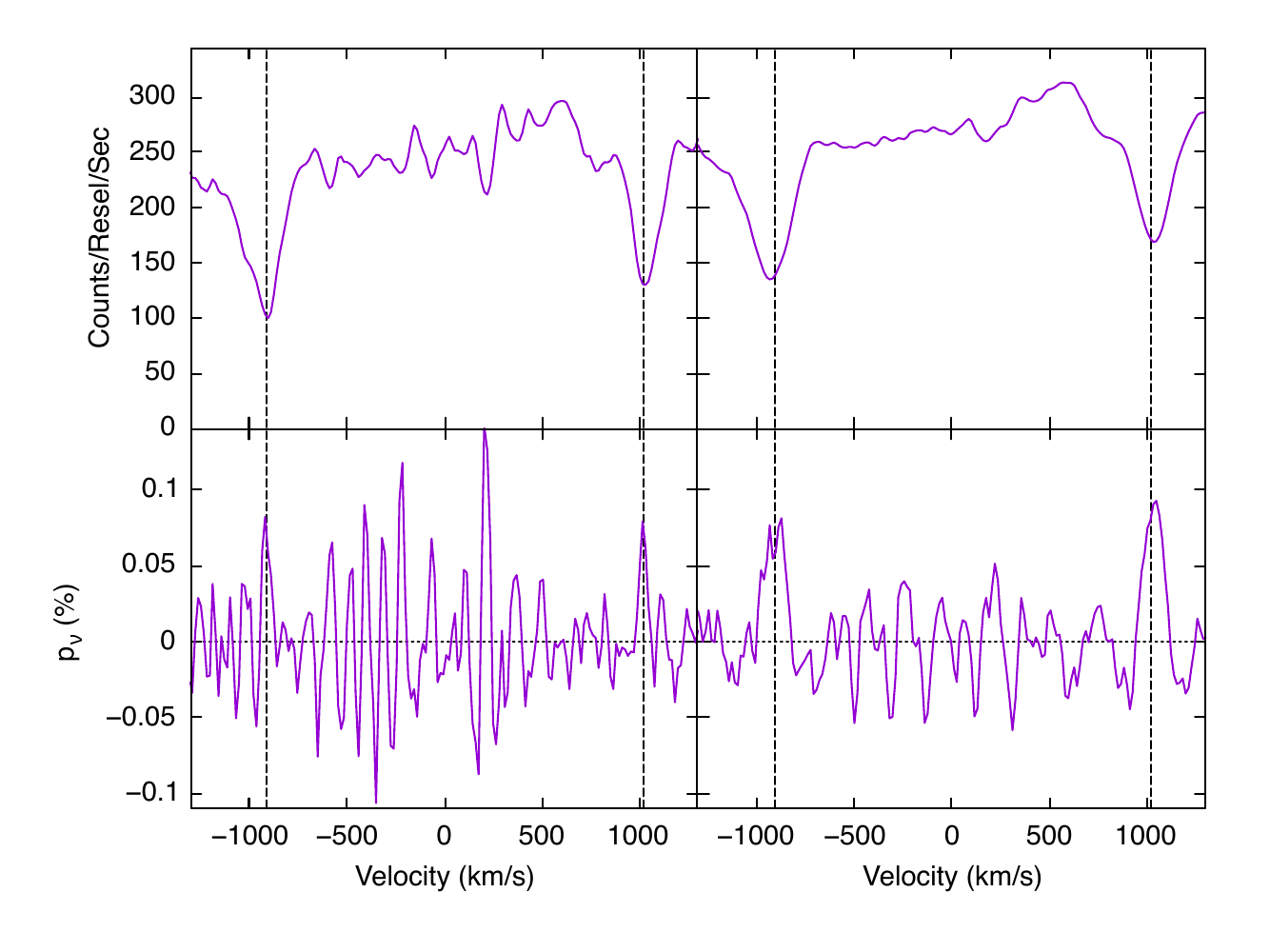}
\caption{Using the simulated spectra of Fig.~\ref{fig18}, shown is a zoomed view around 140~nm for the Si{\sc iv} doublet, indicated by vertical dotted lines.  However, instead of plotting in wavelength, velocity is shown with a zero point at 139.8~nm.} 
\label{fig19}
\end{center}
\end{figure}

Figure~\ref{fig19} shows an expansion of the spectra around the Si{\sc iv} doublet near 140~nm.  Upper is the simulated count rate spectrum, whereas lower is the polarization.  The model is displayed in terms of velocity shift instead of wavelength.  The dotted vertical lines in black indicate line center for each doublet component.

At $v_{\rm rot}=60$ km/s the polarization is quite choppy with some peaks have polarization amplitudes in excess of 0.1\%.  Certainly the doublet components are not the strongest.  At $v_{\rm rot}=120$ km/s the higher blending suppresses the high peaks through polarimetric cancellation resulting in lower-amplitude and less choppy features overall.  However, the Si{\sc iv} are now the dominant features for this slice of the spectrum.  This trend was highlighed in Figure~\ref{fig11} and is preserved in the {\em Polstar} simulation.  

Note that in the case of the longer wavelength component, the signature ``triple peak'' for the \"{O}hman effect survives well for $v_{\rm rot}=120$ km/s (lower right).  Relative to the single trough of the absorption profile in the panel directly above, the polarization feature shows more rapid changes.  This suggests that the polarization profile provides not only an independent measure of the $v \sin i$ for the star, but in some cases possibly a superior measure because the polarized line features are steeper.

\section{Summary}   \label{sec4}

We have presented synthetic spectra in linear polarization based on a grid of plane-parallel stellar atmospheres for hot stars.  The models span temperatures of 15,000~K to 30,000~K with gravities of $\log g$ from 3.0 to 4.0 or in some cases 4.5.  All the models are spherically symmetric, and therefore the continuum polarization cancels identically.  However, rotation leads to a variation in the signed linear polarization across a spectral line.  This variation arises because rotation acts as a absorptive ``mask' that shifts across the face of the star as a function of velocity shift in a spectral line.  This mapping from observer velocity to spatial mask on the star leads to differential absorption that breaks symmetry leading to a net polarization as a function of velocity shift, although the flux of polarization across the line cancels in the net.  This result was first suggested by \cite{1946ApJ...104..460O} for which the effect now bears his name:  a triple-peaked polarized line profile with sign reversals between the peak at line center and the two in the line wings.

FUV wavelengths are emphasized as this is where the effect has greatest amplitude for hot stars, corresponding to spectral regions near the Wien peak of the stars.  The effect is larger simply because the continuum polarization is higher in the FUV, thus any breaking of the spherical symmetry of the atmosphere will yield a polarization that scales with the amplitude of limb polarization.  In some cases the polarization reversal (peak to trough) can be around 1\%.  More typically, it is around 0.01--0.1\%, with higher or lower values achieved depending on wavelength, rotation speed, temperature, and gravity.  In the case of some strong lines, such as Si{\sc iv}, the effect persists even to high rotation speeds at about the 0.1\% level.  

The polarization in lines from 120--160~nm weakens significantly hotter than 30,000~K, because the Wien peak shifts to quite low wavelengths.  This means that even at 120--160~nm, the spectrum is more in the Rayleigh-Jeans tail, lessening the source function gradient, and lower the scale of limb polarization and correspondingly the line polarization amplitude.  Below 15,000~K, the effect can be strong; however, the Wien peak moves longward of the FUV and the flux level drops precipitously, making detection at these wavelengths unlikely for spectral classes cooler than B~type.  Consequently, the \"{O}hman effect is ideally to be detected at FUV wavelengths for B and late O stars.

Numerous lines at these wavelengths lead to lots of signature \"{O}hman profiles, and significant blends depending on the wavelength, temperature, and rotation.  Indeed beyond $v_{\rm rot}$ of about 100 km/s, blending strongly suppresses the polarization amplitude.  We explored an approach that considers the variable polarization statistically in terms of numerous signed polarimetric fluctuations, largest at the shorter wavelengths and declining with increasing wavelength.  Recall that every single absorption produces a triple-peak polarization profile, so 3 peaks over a line FWHM that is around $v_{\rm rot}$.  The distribution of fluctuations, positive and negative, are bell-shaped overall, with significant tails from several stronger lines, but asymmetric.  The widths of these distributions are wider for lower gravities.  Such analysis provides a robust means for understanding the influence of measurement uncertainties for detecting the \"{O}hman effect, both in terms of the statistical fluctuations across a waveband and in terms of how many strong lines could be targeted.

Calculations presented here provide new and novel ways of testing current state-of-the-art stellar atmospheres models, similar to how approaches have been developed for rigorous testing of models with observations of limb darkening \citep[e.g.,][]{2004A&A...413..711W, 2008A&A...482..259C, 2015MNRAS.450.1879E, 2017ApJ...845...65N}. Polarization provides a different dataset for rigorous testing of models by relaxing the assumption of isotropic scattering. The work reported here also supports the effort of probing stellar rotation in massive stars and new mission concepts for undertaking spaceborne UV spectropolarimetry that would be capable of detecting the \"{O}hman effects in hot stars, such as {\em Polstar} or {\em Pollux}.


\appendix 

\section{Polarization from Plane-Parallel Atmospheres with Dipole Scattering}   \label{appA}

The first two Stokes parameters, I and Q, of the radiation emergent from a 
stellar atmosphere can, from the formal solution of the transfer equation, be written (Harrington 1970) as

\begin{eqnarray}
I_\nu(0,\mu)& = & \int_0^\infty\left\{s_\nu(\tau_\nu)+\left(\frac{1}{3} - \mu^2
   \right)p_\nu(\tau_\nu)\right\} \nonumber \\
   & & \times\,  e^{-\tau_\nu/\mu}~\frac{d\tau_\nu}{\mu}
   \label{eq:Inu}
\end{eqnarray}

\begin{equation}
Q_\nu(0,\mu)~=~\int_0^\infty~\left[\left(1 - \mu^2\right)
     p_\nu(\tau_\nu)\right]~e^{-\tau_\nu/\mu}~\frac{d\tau_\nu}{\mu}
     \label{eq:Qnu}
\end{equation}
               
\noindent Here $s_\nu(\tau_\nu)$ is similar to the source function for 
isotropic scattering, while $p_\nu(\tau_\nu)$ is a source term peculair to this
problem. The equations which must be solved to obtain these source functions 
are:

\begin{equation}
s_\nu(\tau_\nu)~=~(1-\lambda_\nu)~\left[\Lambda_{\tau_\nu}(s_\nu)+\frac{1}{3} 
    M_{\tau_\nu}(p_\nu)\right]~+~\lambda_\nu B_\nu(\tau_\nu)
\end{equation}

\begin{equation}
p_\nu(\tau_\nu)~=~\frac{3}{8}(1-\lambda_\nu)~\left[M_{\tau_\nu}(s_\nu)+
          N_{\tau_\nu}(p_\nu)\right] 
\end{equation}
\noindent The quantity $\lambda_\nu$ is the ratio of pure absorption to extinction, and
$(1-\lambda_\nu)$ the ratio of scattering to extinction.
Here, $\Lambda_\tau$ is the familiar $\Lambda$-operator, 

\begin{equation}
\Lambda_\tau\left\{f(t)\right\}~=~\frac{1}{2}\int_0^\infty f(t)~E_1(|t-\tau|)~dt~~~.
\label{eq:Lam}
\end{equation}

\noindent
If we ignore the term involving $M_\tau(p)$, we see that the first equation is 
just the famriliar Schwarzschild-Milne equation for unpolarized radiation.
The new $M_\tau$ and $N_\tau$ operators (which have no relation to the 
$M_n(\tau)$ and $N_n(\tau)$ used in Kourganoff 1963) are defined as

\begin{equation}
M_\tau\left\{f(t)\right\}=\int_0^\infty f(t)\left[\frac{1}{2}E_1(|t-\tau|)
           - \frac{3}{2}E_3(|t-\tau|)\right]~dt
           \label{eq:M}
\end{equation}

\begin{eqnarray}
N_\tau\left\{f(t)\right\} & = & \int_0^\infty f(t)\left[\frac{5}{3}E_1(|t-\tau|) -4 E_3(|t-\tau|) \right.  \nonumber \\ 
   & + &  \left.  3 E_5(|t-\tau|)\right]~dt
   \label{eq:N}
\end{eqnarray}

\noindent We solve these equations by first adopting a grid of optical depth points  
$\tau$: $\tau_1,\tau_2, ..., \tau_N$. The solution will be represented by 
 two vectors, $\vec{s} = s_1, s_2, ... , s_N$ and $\vec{p} = p_1, p_2, ... , p_N$. We will represent the $\Lambda$-, $M$-, amd $N$- operators of equations (\ref{eq:Lam}),
(\ref{eq:M}) and (\ref{eq:N}) by $N\times N$ matrices. E.g., we will find a matrix $\Lambda_{ij}$ 
such that for a vector $\vec{f} = f_1, f_2, ... , f_N$ of the values of some
function $f(\tau)$ at our $\tau$-points, the matrix product $\Lambda_{ij} \vec{f}$ gives the $\Lambda$ transform of the function $f$ at the $\tau$ points. 

We assume that $s$ and $p$ can be well approximated by cubic spline functions. The key point is that the defining equations for the cubic splines (e.g., the continuity of the second derivatives) can be written as a matrix of coeficients times
the vector of unknowns where the elements of the coefficient matrix are 
functions only the $\tau$ vector.  Further, the operators $\Lambda_\tau$,
$M_\tau$ and $N_\tau$ operating on the splines can be evaluated by analytic integrals of the form

\begin{equation}
\int_{\tau_i}^{\tau_{i+1}} \tau^n~\Lambda_\tau~d\tau~~~~\mbox{for  n=0,1,2,3}, 
\end{equation}
\noindent and similar integrals for $M_\tau$ and $N_\tau$. 

We thus obtain the following system of linear equations for $s_i$ and $p_i$, where 
the left-hand side involves pre-computed matrices.\footnote{For more detailed expressions, see www.astro.umd.edu/$\sim$jph/Matrix\_Op\_Pol}


\begin{equation*} {\left[ \begin{array}{*{2}{c}}
  I_{ij}-(1-\lambda_i)\Lambda_{ij} &~~~-\frac{1}{3}(1-\lambda_i)M_{ij}
  \\ -\frac{3}{8}(1-\lambda_i)M_{ij}
   & ~~I_{ij}-\frac{3}{8}(1-\lambda_i)N_{ij} \end{array} \right] }~~
   {\left[ \begin{array}{*{1}{c}} s_i \\ p_i \end{array} \right] }~~
\end{equation*}
\begin{equation}
  \hspace{1.5in} ={\left[ \begin{array}{*{1}{c}} \lambda_i~B_i \\ 0 \end{array} \right] }
\end{equation}

\noindent where $I_{ij}$ is the identity matrix and $B_i = B_\nu[T(\tau_i)]~$. 
 Note that $\lambda_i~$, the 
fraction of the opacity due to pure absorption, may vary with $\tau_i$.

This linear system may be solved 
by any standard method. Since $\lambda_i~B_i$ will differ for each
wavelength, it is best to compute one set of operator matrices for a standard 
$\tau_i$ grid and interpolate $\lambda_i$ and $B_i$ to that specific $\tau$ scale.

Finally, we use a similar approach to extracting the Stokes parameters $I$ and $Q$, 
which are a function of $\mu=\cos(\theta)$. We choose a discrete set of $\mu$'s,
sufficient for interpolation (we use at least 18). The operator in this case is
\begin{equation}
E_\mu\{f(t)\}~=~\int_0^\infty f(t)~e^{-\tau/\mu} \frac{d\tau}{\mu},
\end{equation}

\noindent where $f(t)=s(t)$ or $p(t)$.

By the same spline approximation, we can obtain a matrix operator $E_{m,n}$.
If $\vec{\mu}=\mu_1,\mu_2, ... \mu_M$, then we have an M x N matrix.
From equations (\ref{eq:Inu}) and (\ref{eq:Qnu}), the emergent radiation intensity $I$ and the Stokes
$Q$ are given by

\begin{equation}
 I(0,\mu_m)=\sum_{n=1}^N E_{m,n}~s_n + \left(\frac{1}{3}-\mu^2 \right) 
    \sum_{n=1}^N E_{m,n}~p_n  
\end{equation}
\begin{equation}
  Q(0,\mu_m)= (1 - \mu^2)~\sum_{n=1}^N E_{m,n}~p_n
\end{equation}

\noindent We have used this method for all the results reported here.

\section{A Grid of Second Spectra for Hot Stars}    \label{appB}

\begin{figure*}
\begin{center}
\includegraphics[width=0.48\textwidth]{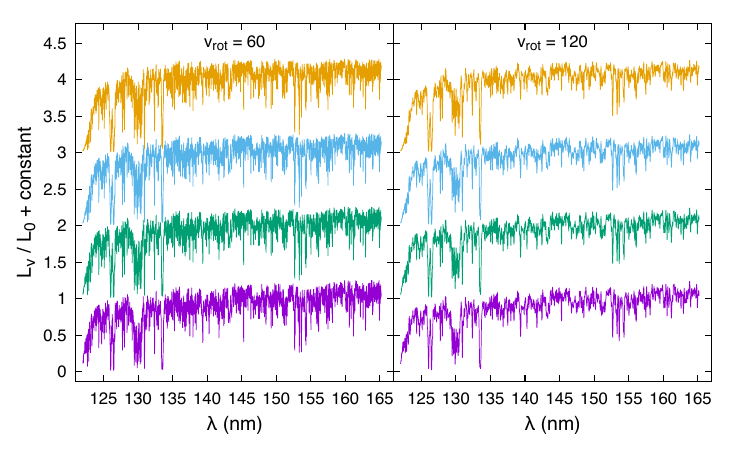}
\includegraphics[width=0.48\textwidth]{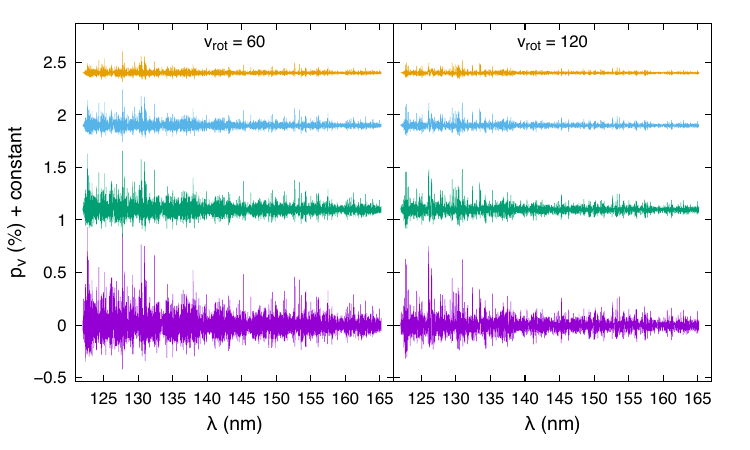}
\caption{Plane-parallel stellar atmosphere models with $T_{\rm eff} = 15,000$~K.  Left and right pairs are for the total light spectra and the polarized spectra, respectively.  Models for rotation speed with $v_{\rm rot}=60$ and 120 km/s were calculated, as labeled.  Each panel has 4 colors for $\log g = 3.0$ (purple), 3.5 (green), 4.0 (blue), and 4.5 (yellow).} 
\label{fig12}
\end{center}
\end{figure*}

\begin{figure*}
\begin{center}
\includegraphics[width=0.48\textwidth]{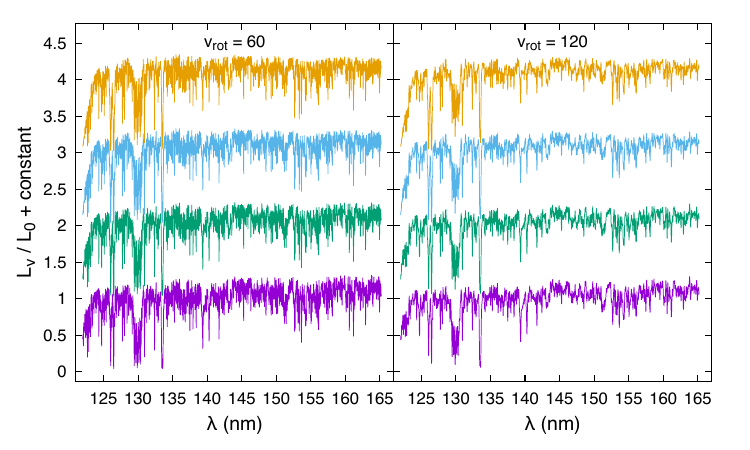}
\includegraphics[width=0.48\textwidth]{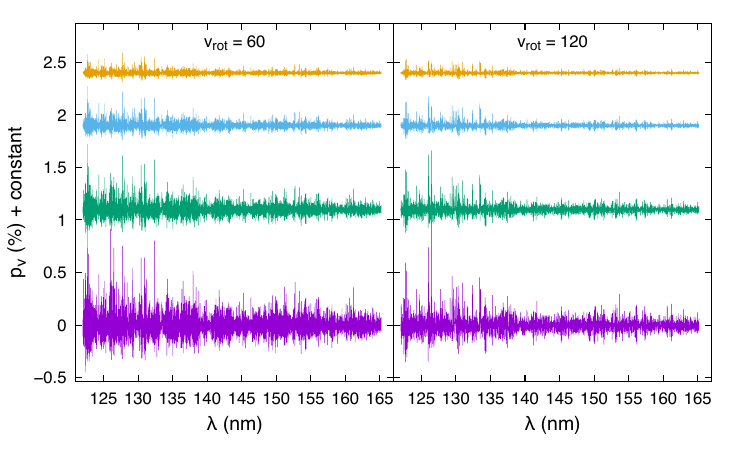}
\caption{As Fig.~\ref{fig13} for $T_{\rm eff}=17,000$~K.} 
\label{fig13}
\end{center}
\end{figure*}

\begin{figure*}
\begin{center}
\includegraphics[width=0.48\textwidth]{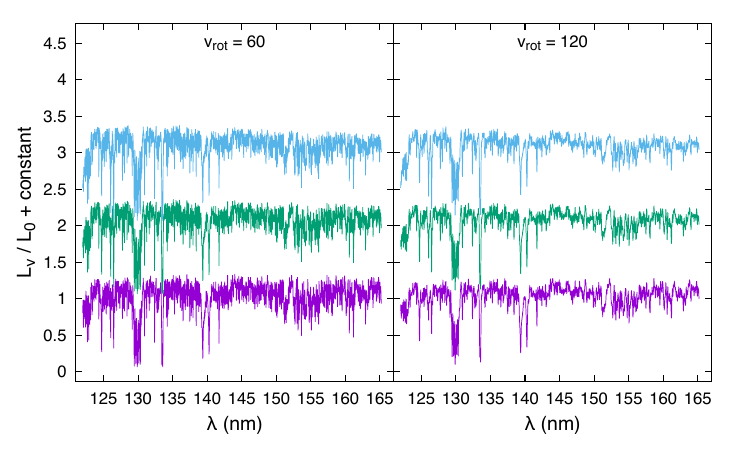}
\includegraphics[width=0.48\textwidth]{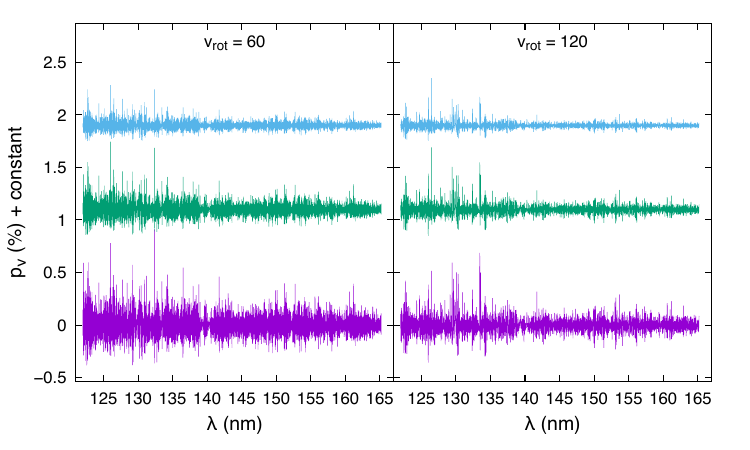}
\caption{As Fig.~\ref{fig13} for $T_{\rm eff}=20,000$~K, but without $\log g = 4.5$.} 
\label{fig14}
\end{center}
\end{figure*}

\begin{figure*}
\begin{center}
\includegraphics[width=0.48\textwidth]{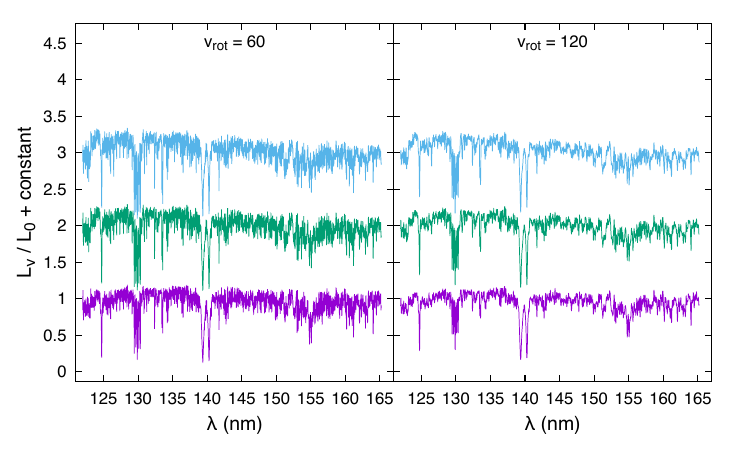}
\includegraphics[width=0.48\textwidth]{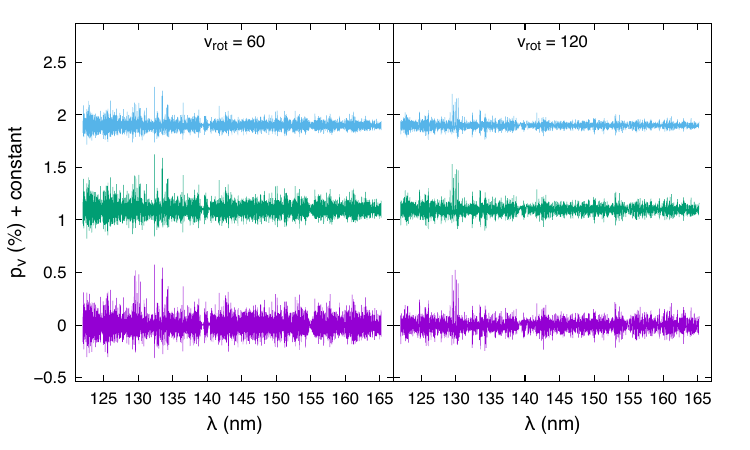}
\caption{As Fig.~\ref{fig13} for $T_{\rm eff}=25,000$~K, but without $\log g = 4.5$.} 
\label{fig15}
\end{center}
\end{figure*}

\begin{figure*}
\begin{center}
\includegraphics[width=0.48\textwidth]{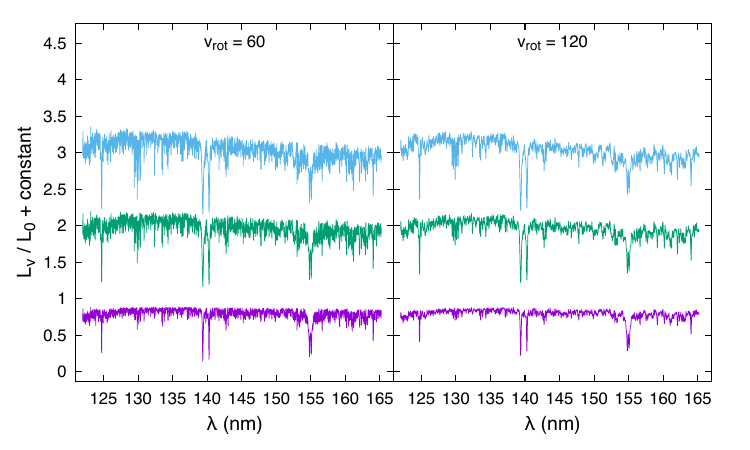}
\includegraphics[width=0.48\textwidth]{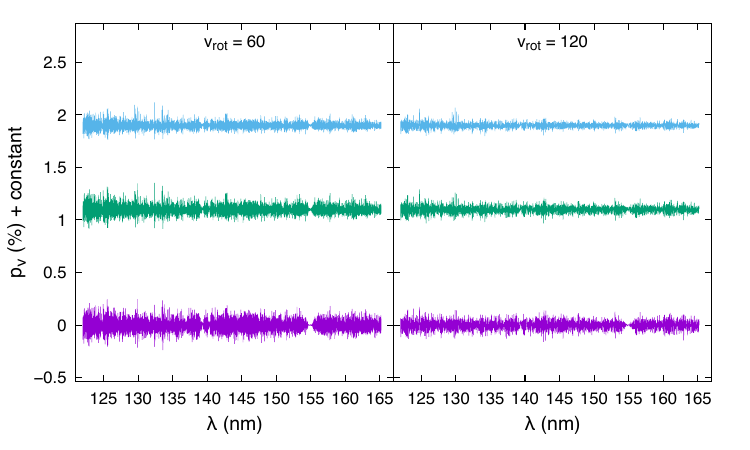}
\caption{As Fig.~\ref{fig13} for $T_{\rm eff}=30,000$~K, but without $\log g = 4.5$.} 
\label{fig16}
\end{center}
\end{figure*}

Figures~\ref{fig12}--\ref{fig16} display results for spherical stars based on plane-parallel model atmospheres methods detailed in App.~\ref{appA}.  The left pair panels are for the total light spectra; the right pair panels are for the polarized spectra.  Models with rotational speeds of $v_{\rm rot}=60$ km/s and 120 km/s were calculated, as labeled.

The spectra focus on FUV wavelengths from about 120~nm to 165~nm.  On the selection of the short wavelength, some instrument designs make use of MgF$_2$ coatings for measurement of polarization, which cuts off around 121~nm \citep[e.g.,][]{2015ApOpt..54.7377P}.  For the selection of the long wavelength, as the Figures themselves attest, the polarization amplitude of the \"{O}hman effect is dropping with increasing wavelength, and simply becomes smaller beyond 165~nm, because of the lessened source function gradient that controls the polarization as the spectra shift more into the Rayleigh-Jeans tail. 

The stellar temperatures are, in order of the figures, $T_{\rm eff} = 15,000$, 17,000, 20,000, 25,000, and 30,000~K.  For the selection of temperatures, by 40,000~K, the polarization amplitudes of lines are small because the Wien peak has moved shortward of 120~nm.  We chose 30,000~K as our hottest temperature as representative of early B stars and late O stars.  At cooler temperatures the polarization can still be large, but the Wien peak shifts toward longer wavelengths.  We chose 15,000~K as our coolest value and representative of late B~stars.

Each figure has multiple curves.  For 15,000~K and 17,000~K models, the 4 curves are for gravities $\log g = 3.0$ (purple), 3.5 (green), 4.0 (blue), and 4.5 (yellow).  For the higher temperatures, $\log g=4.5$ is omitted because such high gravities tend to be stars below the main sequence and because the polarization values tend to be quite low.

Each figure is for a different temperature, and each has four panels.  In each Figure, the two at left are for $v_{\rm rot}=60$ km/s and the two at right are for $v_{\rm rot}=120$ km/s.  Top panels are relative scaled fluxes; bottom are for signed polarization, $q=100\% \times L^Q/L^I$.  For 15,000 and 17,000~K models, results are shown for $\log g =3.0$, 3.5, 4.0, and 4.5.  For 20,000, 25,000, and 30,000~K, we omit $\log g=4.5$ since early B stars on the main sequence tend to have gravities closer to $\log g \sim 4$.  Moreover, the polarization amplitudes of lines are quite small.

\section*{Acknowledgements}

We are grateful to the referee for several comments that
have improved this manuscript.

\section*{Funding Statement}

The authors declare that no funds, grants, or other support were received during the preparation of this manuscript.

\section*{Ethics Approval}

Not applicable

\bibliography{ohman}

\end{document}